\documentclass[journal,onecolumn]{IEEEtran}
\usepackage{comment} 
\usepackage{amsmath}  
\usepackage{float}

\usepackage{ifpdf}
\usepackage{cite}
 
\ifCLASSINFOpdf
  \usepackage[pdftex]{graphicx}
\else
  \usepackage[dvips]{graphicx}
\fi
\usepackage{algorithmic}
\usepackage[utf8]{inputenc}
\usepackage[english]{babel}
\usepackage{amsthm}
\usepackage{amsfonts}
\usepackage{array}
\usepackage{mdwmath}
\usepackage{mdwtab}
\usepackage{graphicx} 
\usepackage{caption}  
\usepackage{amssymb}
\usepackage{subcaption}
\usepackage{booktabs}
\usepackage[ruled]{algorithm2e} 
\usepackage{url}

\newcommand{\squeezeup}{\vspace{-3mm}}
\newcommand{\nn}{\nonumber \\}
\allowdisplaybreaks

\ifCLASSINFOpdf
\else
\fi
\begin{document}
%
\title{Platoon Stability and Safety Analysis of Cooperative Adaptive Cruise Control under Wireless Rician Fading Channels and Jamming Attacks}
%
%
%
\author{\IEEEauthorblockN{Amir Alipour-Fanid, Monireh Dabaghchian and Kai Zeng}\\
\IEEEauthorblockA{Electrical and Computer Engineering Department}, George Mason University, Fairfax, Virginia 22030, USA\\
Email: \{aalipour, mdabaghc, kzeng2\}@gmu.edu}

\maketitle

\begin{abstract}
Cooperative Adaptive Cruise Control (CACC) is considered as a key enabling technology to automatically regulate the inter-vehicle distances in a vehicle platoon to improve traffic efficiency while maintaining safety.
Although the wireless communication and physical processes in the existing CACC systems are integrated in one control framework, the coupling between wireless communication reliability and system states is not well modeled. 
Furthermore, the research on the impact of jamming attacks on the system stability and safety is largely open.
In this paper, we conduct a comprehensive analysis on the stability and safety of the platoon under the wireless Rician fading channel model and jamming attacks.
The effect of Rician fading and jamming on the communication reliability is incorporated in the modeling of string dynamics such that it captures its state dependency. 
Time-domain definition of string stability is utilized to delineate the impact of Rician fading and jamming on the CACC system's functionality and string stability.
Attacker's possible locations at which it can destabilize the string is further studied based on the proposed model.
From the safety perspective, reachable states (i.e., inter-vehicle distances) of the CACC system under unreliable wireless fading channels and jamming attacks is studied.
Safety verification is investigated by examining the inter-vehicle distance trajectories. 
We propose a methodology to compute the upper and lower bounds of the trajectories of inter-vehicle distances between the lead vehicle and its follower. 
We conduct extensive simulations to evaluate the system stability and safety under jamming attacks in different scenarios.
We identify that channel fading can degrade the performance of the CACC system, and the platoon's safety is highly sensitive to jamming attacks.
The best location to launch the jamming attack to destabilize the platoon is above the second vehicle in the platoon.
The platoon is more vulnerable to jamming attacks when the lead vehicle is decelerating.\end{abstract}

\begin{IEEEkeywords}
Cooperative Adaptive Cruise Control, Vehicle-to-Vehicle Communication, Wireless Rician Fading Channel,
Jamming Attacks, Platoon Stability, Safety, Reachability Analysis
\end{IEEEkeywords}

%
\IEEEpeerreviewmaketitle

\ifCLASSOPTIONcaptionsoff
  \newpage
\fi



%

\section{Introduction}
\label{Introduction}
\IEEEPARstart{V}{ehicular} Cyber-physical systems (CPS) expand the capabilities of the vehicles through the integration of computation, communication, and control \cite{DC2015}.
Vehicle platooning is one of the important vehicular CPS applications that operates based on tight coupling of cyber part (wireless communication) and physical processes (Vehicles dynamics response and inter-vehicle distances). 
It will be an indispensable part of intelligent transportation system (ITS) in the future \cite{HB2014}.  

Cooperative Adaptive Cruise Control (CACC) system as an extension of Adaptive Cruise Control (ACC) is proposed to improve vehicle platooning performance and efficiency \cite{SN2014, GR2010,SJ2014}. 
With CACC, vehicles in a platoon adjust their inter-vehicle distances autonomously such that they can line up as close as possible in order to improve traffic throughput.
In the CACC system, absolute relative distance and velocity information is measured by the radar and preceding vehicle's acceleration information is sent over the wireless link to the follower vehicle.
These information are fed into the feedback and feedforward controllers to compute the control command of the corresponding vehicle. 
CACC system alleviates traffic congestion, improves mobility and increases road safety \cite{BC2006}. 
In addition, this technology reduces fuel consumption and provides better comfortability for the passengers compared with solely human controlled vehicles \cite{Lang2014}.  

Despite the tremendous benefits provided by CACC system, the wireless communication between vehicles is subject to channel fading and jamming attacks, which leads to undesirable packet loss, and in turn introduce significant disturbances in safe and efficient operation of the system \cite{MA2015}.
However, the impact of wireless channel fading and jamming attacks on the stability and safety of CACC systems is not well modeled nor understood in the existing literature.
The main challenge lies in the tight coupling of cyber (wireless communication) and physical states (inter-vehicle distance).
In a CACC enabled vehicle platoon, the distance between vehicles may change depending on the lead vehicle's behavior and spacing policy \cite{VS2014}. 
This variation in inter-vehicle distance affects the wireless channel conditions in terms of fading and path loss, which further affects the received-signal-strength (RSS) and packet delivery ratio. 
However, in the existing literature, the consideration of this coupling between the system state (inter-vehicle distance) and wireless channel conditions is missing \cite{GR2010, SJ2014, XA2001, JD2014, QG2014}.



In this paper, we investigate the performance of CACC system subject to channel fading and jamming attacks.
Dynamic inter-vehicle distance variation will change the instantaneous channel quality condition established between two consecutive vehicles in the platoon. 
On the one hand, wireless channel state condition has a direct and effective impact on the inter-vehicle distance state evolution in the vehicle longitudinal modeling. 
On the other hand, the inter-vehicle distance state evolution will have a strong effect on channel state condition as well.
This cyber and physical state coupling is addressed in this paper which plays an important role in analyzing CACC system's stability and safety.
We also need to mention that the words \emph{platoon} and \emph{string} are used interchangeably in this paper.

Furthermore, with the assumption of Rician fading channel, we study the string stability of the CACC system under a mobile jamming attack. 
The attacker jams the wireless channel established among the vehicles in order to prevent the receivers from decoding the packets with the purpose of destabilizing the platoon. 
If the attacker is successful in jamming the wireless communication, the CACC system will not work in the normal condition until the next packet is received successfully. 
We evaluate the impact of jamming attack on the mean string stability by employing the time-domain definition of string stability.

We summarize the main contributions of the this paper as below. 

\begin{itemize}
\item In this work, the cyber-physical state coupling in platoon with CACC is modeled. That is, the coupling between cyber (wireless communication) and physical state (inter-vehicle distance) is addressed.
\item Minimum headway-time value is obtained in the frequency domain by non-liner programming. The time-domain string stability is validated by the frequency domain analysis.  
\item Based on the modeling, we analyze the platoon stability considering various scenarios with different settings, including various attacker's and vehicle's signal transmission power. 
As a defending strategy against jamming attack, minimum vehicle's signal transmission power is computed such that the mean string stability is maintained.  
\item We study the attacker's best locations at which it can destabilize the string when the communication channels are under jamming attacks. 
\item We study the impact of Rician fading on the reachable inter-vehicle distance states in the platoon. 
\item We propose a methodology to compute the lower and upper bounds of the inter-vehicle distance between the lead vehicle and its follower. The impact of Rician fading and jamming attacks on the inter-vehicle distance trajectories are studied from the safety perspective. 
\item We conduct extensive simulations to analyze the string stability and the inter-vehicle distance states evolution in the platoon under various system settings and scenarios.  
\end{itemize}
 


Through our analysis and simulation evaluation, we make the following findings.

\begin{itemize}
\item Channel fading can degrade the performance of the CACC system. 
\item The platoon is highly sensitive to jamming attacks, and its stability and safety can be compromised by a jammer.
\item The attacker's location being close to the second vehicle in the platoon (first vehicle following the lead vehicle) is the best location for the mobile jamming attacker to destabilize the platoon.
\item The platoon is more vulnerable to jamming attacks when the lead vehicle is decelerating. 
This finding is expected to motivate more future research on physical state-aware cyber attacks and defenses for CACC systems in specific and cyber-physical systems in general.
\end{itemize}
 

\section{Related Work}
\label{RelatedWork}  

String stability has been analyzed in the frequency domain for the mass-spring-damper framework of unidirectional (forward-looking) and bidirectional (forward-and-backward-looking) control strategies by Diana et al. \cite{YK1996}.  
Required conditions for control parameters and variable headway-time have been derived for the constant and velocity-dependent space policy.
Inter-vehicle communication is also integrated for the constant spacing policy with unidirectional control scheme. 
Vehicles in the platoon receive the lead vehicle's velocity information through wireless communication. 
This information flow is modeled as an additional damper, connecting each vehicle's mass in the platoon to the lead vehicle's mass.
This work represents a fundamental analysis on string stability for different control strategies.
However, in the cooperative (forward-looking with wireless communication) framework, it does not address the impact of wireless channels' uncertainty introduced by path loss and fading on the system performance.
Wireless jamming attack is not a concern in this work. 

Similarly, other existing works \cite{GR2010, PS2011, EJ2007} consider normal operation of CACC system without any consideration of packet loss due to wireless channel fading or jamming attacks.
The frequency response of the system is derived in these cases and the string stability is analyzed in a fairly nice format in the frequency-domain. 

Necessary and sufficient conditions for string stability of a heterogeneous platoon are studied by Naus et al. \cite{GR2010}.
Network delay and sampling effects are introduced in the string stability analysis by \"Onc\"u et al. \cite{SJ2014}.
The delay is assumed identical in all the communication links and string stability is investigated for various sampling intervals and headway-times.
In \cite{XA2001}, the robustness of a CACC system to communication delays is studied and an upper bound on the delay required for stability is derived.
However, the impact of inter-vehicle distance on the wireless communication reliability is not considered in these works.   

There are few works studying the security of vehicle platooning in terms of attacking on wireless communication or control components. 
In \cite{SR2015}, an insider attacker attacks on controller gains of a vehicle in the platoon.
The attacker has the capability of modifying the gains such that it can destabilize the platoon. 
In \cite{GR2013}, mass-spring-damper follower dynamics model is considered for studying the platoon performance under attack.
A new class of the attack based on vehicle misbehavior is proposed.
It shows that the attacker is effective when the attacker is near the rear of the platoon. 
However, our work is different from \cite{GR2013} in terms of platoon modeling, attacker's nature, purpose of the attack, and the evaluation method employed to measure the impact of the attack. 
In another work \cite{MA2015}, various security vulnerabilities on the CACC system have been identified. 
Message falsification and radio jamming attack's effect are studied through Vehicular Network Open Simulator. 
However, the CACC control structure and jamming attack strategies are considered as a black box in the simulation environments.
The coupling between the system states and wireless communication channel condition is not well modeled.

In \cite{QG2017}, string stability under stochastic communication delays has been studied.
Packet drops introduce random delay that follows geometric distribution for each discrete time.  
Non-linear controller's gains are designed such that the string stability can be maintained while wireless communication channels among vehicles suffer from packet loss.    
In this work, it is assumed that packet drop distribution is independent of string's dynamics instant states.
In other words, state dependency of packet loss has been ignored.
However, as we identified, physical states variation influence communication reliability and vise verse.   
Path loss, fading impact on packet delivery ratio, and system's performance are missing in this work. 

In our previous work \cite{AD2017}, we considered two-ray ground-reflected propagation model on the channels \cite{CS2012}. 
However, in the current paper, Rician fading channel is modeled such that it takes the distance state dependency into consideration. 
In \cite{AD2017}, string stability has been analyzed based on fixed setting parameters, while in this paper, we examine various system parameters and settings.
Furthermore, we investigate the impact of jammer's and vehicle's signal power on the string stability.
We propose a methodology to compute the lower and upper bounds of the inter-vehicle distance between the lead vehicle and its follower.
The impact of Rician fading and jamming attacks on the inter-vehicle distance trajectories are studied from the safety perspective.   
  
\section{System Model}
\label{SystemModel}
 
In this section, we describe the models of platoon with CACC, wireless channel, and a mobile jammer. 

\subsection{Vehicle String}
\begin{figure*}[ht!]
						  \centering
								\includegraphics[scale=.5]{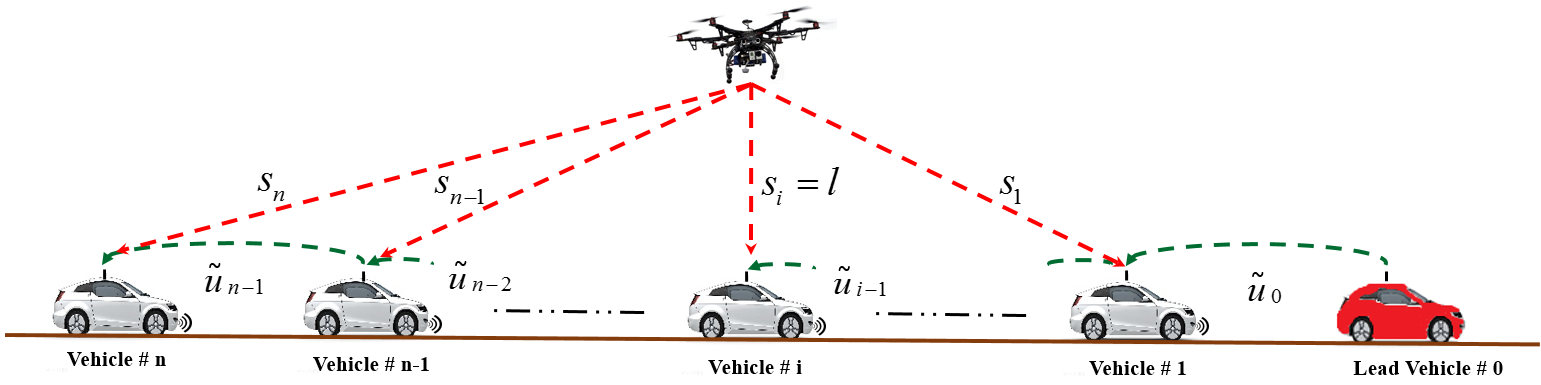}
							\caption{Vehicle platoon under Rician fading channel and mobile jamming attacker}
						  \squeezeup
							\label{platoon}
						\end{figure*}

We consider a platoon of vehicles consisting of ($n+1$) homogenous vehicles (identical longitudinal dynamic properties) shown in Figure~\ref{platoon}. 
Each vehicle is equipped with a CACC system. 
In other words, each vehicle is equipped with a radar in front of the vehicle to measure the absolute relative distance from the vehicle ahead of it.
At the same time, V2V wireless communication (e.g., using IEEE $802.11$p Dedicated Short Range Communication (DSRC) technology \cite{Li2012}) is used to transmit each vehicle's acceleration information to its following vehicle.
Similar to the assumptions in \cite{SN2014}, the acceleration information of each vehicle is sent every $100$ms to the following vehicle. 

\subsection{Wireless Channel}
Each vehicle in the platoon receives/sends the acceleration information from/to its immediately preceding/following vehicle.
Therefore, a wireless channel is established between each two consecutive vehicles in the platoon.  

In the vehicle platooning with CACC and velocity-dependent spacing policy, the inter-vehicle distances are changing according to the control parameters' setting and the lead vehicle's action.
Thus, because of the vehicles' mobility and possible line-of-sight and multipath signal propagation, we assume Rician fading channel model for the wireless communication channels among the vehicles.
In Rician fading stochastic model, direct path signal's power appears stronger than the signal's power in the scattered paths.
However, the received signal amplitude is subject to fading which follows the Rician fading distribution \cite{WL2009}. 
Moreover, depending on the inter-vehicle distances, free space path loss affects the received signal power level at the receiver of each vehicle.


\subsection{Attacker}
We consider a mobile jammer, which can be mounted on a drone flying over the platoon.  
Since the power source of the drone is limited, we assume a reactive jammer \cite{WW2005}.
A reactive jammer has the capability of sensing channels and launching its jamming signal whenever the vehicles transmit their acceleration information through the wireless medium \cite{WW2005}.
All the legitimate established wireless links among each pair of transmitters and receivers in the platoon are under jamming attacks.




\section{CACC Control Structure and String State Space Representation}  
\label{CACCstructureSSR}  
\subsection{Longitudinal Vehicle Dynamics}
The common linearized third-order state space representation used for modeling longitudinal vehicle dynamics is as follows \cite{SJ2014} 
\begin{equation}
\begin{aligned}
\label{equ1}
\dot q_i(t) = v_i(t), \quad \dot v_i(t)= a_i(t), \quad \dot a_i(t)=-\eta_i^{-1}+\eta_i^{-1}u_i(t), \quad \text{for} \quad  i=0,1,...,n
\end{aligned}
\end{equation}
where $q_i(t)$, $v_i(t)$, and $a_i(t)$ are absolute position, velocity, and acceleration of the $i$th vehicle, respectively.
$\eta_i$ and $u_i(t)$ represent the internal actuator dynamics and the commanded acceleration of the $i$th vehicle, respectively.
The transfer function of the longitudinal vehicle dynamics $G_i(s)$ is derived as follows:
\begin{equation}
\begin{aligned}
\label{equ2}
G_i(s) = \frac{Q_i(s)}{U_i(s)}= \frac{1}{s^2(\eta_i s + 1)} \quad \text{for} \quad  i=0,1,...,n
\end{aligned}
\end{equation}
where $Q_i(s) = \mathcal{L}(q_i(t)) $ and $U_i(s) = \mathcal{L}(u_i(t)) $ represent the Laplace transformation of the absolute position and the commanded acceleration of the $i$th vehicle, respectively. 
\subsection{CACC Control Structure}

\begin{figure}[t]
					  \centering 
						\includegraphics[scale=.34]{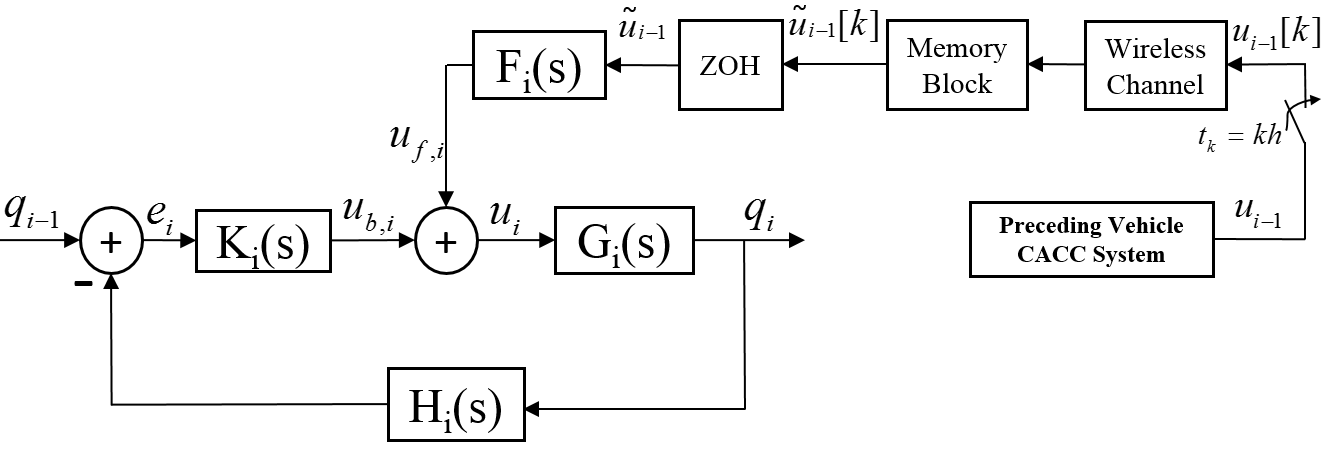}
							\caption{CACC control structure}
							\squeezeup
							\label{CACCmodelattack}  
\end{figure}

The structure of a CACC system is shown in Figure~\ref{CACCmodelattack}.
In this model, $H_i(s)=1+h_ds$ represents the spacing policy dynamics. 
Headway-time constant, $h_d$, indicates the time that it takes vehicle ($i-1$) to arrive at the same position as its preceding vehicle ($i$). 
Several spacing policies have been studied in the literature \cite{YK1996, GR2010}.
In this paper, we consider velocity-dependent spacing policy for the control structure of the CACC system which has been used in \cite{GR2010,SJ2014,YK1996} . 
This spacing policy assists each vehicle in the platoon to not only maintain a safe distance with its preceding vehicle at high speeds, but also increases the traffic throughput on the roads by reducing the inter-vehicle distances as much as possible.
String stability requirement is highly affected by the value of headway-time $h_d$; as a result, this parameter plays a crucial role in operating a safe and efficient CACC system.
Considering velocity-dependent spacing policy, the desired distance is defined as $h_d v_i(t)$.
That is, the distance between the two vehicles increases if the velocity of the preceding vehicle increases, and vice versa. 
Therefore, spacing error $e_i(t)$, at each time instant $t$ can be determined by the difference between the actual relative distance, $d_i(t) = q_{i-1}(t)-q_i(t)$, measured by the radar, and the desired distance, $h_d v_i(t)$, as follows:
\begin{equation}
\begin{aligned}
\label{equ3}
e_i(t) = q_{i-1}(t)-q_i(t)-h_dv_i(t)
\end{aligned}
\end{equation}
In the CACC control structure shown in Figure~\ref{CACCmodelattack}, $K_i(s)=k_{p,i}+k_{d,i}s$ is a feedback (PD) controller where $k_{d,i}$ is the bandwidth of the controller and is chosen such that $k_{d,i}<<1/\eta_{i}$ \cite{SJ2014}. 
The PD controller parameters $k_{p,i}$ and $k_{d,i}$ are set up in such a way that the internal stability of the vehicle dynamics is satisfied. 
In \cite{GR2010}, the feedforward controller $F_i(s)=(H_i(s)G_i(s)s^2)^{-1}$ has been designed such that the zero steady state spacing error ($e_i(t)=0$ as $t\rightarrow \infty$) defined in (\ref{equ3}) is achievable.
In the CACC control structure, $u_{b,i}$ and $u_{f,i}$ represent the feedback and feedforward controllers' output, respectively. 
The summation of these two outputs provide the commanded acceleration $u_i$ for the $i$th vehicle. 

We also use a low cost memory block in the CACC control structure which has the capacity for saving only one packet information. 
Each time, if the memory receives the packet successfully, it updates the information; otherwise, it keeps the last successful received information.
This policy will be modeled in section V and then incorporated in state space representation of the string under wireless Rician fading channels and jamming attacks.  
The ZOH (Zero Order Holder) in the CACC control structure also converts the input discrete-time signal to the continuous-time signal which then is fed into the feedforward controller $F_i(s)$ shown in Figure~\ref{CACCmodelattack}.
 
\subsection{CACC State Space Representation}

State space representation of the CACC control structure in Figure~\ref{CACCmodelattack} is given in \cite{SJ2014}. 
For further analysis we add the inter-vehicle distance state to this representation as follows: 
\begin{equation}
\begin{aligned}
\label{equ4}
\dot d_i(t) &=v_{i-1}(t)-v_i(t)\\
\dot e_i(t) &=v_{i-1}(t)-v_i(t)-h_da_i(t)\\
\dot v_i(t) &=a_i(t)\\
\dot a_i(t) &=-\eta_i^{-1}a_i(t)+\eta_i^{-1}u_i(t)\\ 
\dot u_{f,i}(t)&=-h_d^{-1}u_{f,i}(t)+h_d^{-1}\tilde{u}_{i-1}(t)\\
\end{aligned}
\end{equation}
The commanded acceleration of the $(i-1)$th vehicle, $u_{i-1}(t)$, is transmitted through the wireless channel to the $i$th vehicle. 
The received acceleration information is denoted by $\tilde{u}_{i-1}(t)$ at the receiver of the following vehicle $i$.
From~(\ref{equ4}) we see that the output of the feedforward controller, $u_{f,i}(t)$, depends on the received commanded acceleration, $\tilde{u}_{i-1}(t)$, of the $(i-1)$th vehicle. 
For simplicity, we omit the continuous-time domain representation $t$ in the remained article. 
The commanded acceleration $u_i$, which is the summation of feedback and feedforward controller' outputs, is derived as follows:
\begin{equation}
\begin{aligned}
\label{equ5}
u_i    &= u_{b,i} + u_{f,i}\\ 
       &= k_{p,i} e_i + k_{d,i} \dot e_i + u_{f,i}\\
			 &= k_{p,i} (d_i-h_d v_i) + k_{d,i} (v_{i-1}-v_i-h_d a_i) + u_{f,i}\\
\end{aligned}
\end{equation}
By substituting~(\ref{equ5}) in~(\ref{equ4}), the continuous-time CACC state space representation can be expressed as
\begin{equation}
\begin{aligned}
\label{equ6}
\dot x_i = A_{i,i} x_i + A_{i,i-1} x_{i-1} + B_c \tilde{u}_{i-1}
\end{aligned}
\end{equation}
where $x_i^T=[d_i \quad e_i \quad v_i \quad a_i \quad u_{f,i}]$, for $i=1,2,...,n$ and

\begin{centering}
\quad \quad \quad \quad \quad \quad \quad \quad \quad \quad \quad $A_{i,i}=$$\begin{bmatrix}
      0 & 0 & -1  &      0      &       0        \\
			0 &  0 & -1  &      -h_d      &       0        \\
     0 &  0 &  0  &      1       &       0         \\
     0 &  -\eta_i^{-1} k_{p,i} &  -\eta_i^{-1}(k_{p,i}h_d + k_{d,i})  & -\eta_i^{-1}(1+ k_{d,i} h_d) &      \eta_i^{-1}         \\
			0 & 0 &  0  &      0       & -h_d^{-1}        
     \end{bmatrix},$
\end{centering}\\

\begin{centering}
\quad \quad \quad \quad \quad \quad \quad \quad \quad \quad \quad \quad \quad \quad \quad \quad $A_{i,i-1}=$$\begin{bmatrix} 
    0 &   0 &  1  & 0 & 0       \\
		0 &	 0 &  1  & 0 & 0       \\
    0 &   0 &  0  & 0 & 0       \\
    0 &   0 &  \eta_{i}^{-1}k_{d,i}  & 0 & 0       \\
		0 &	 0 &  0  & 0 & 0        
     \end{bmatrix},$
\quad		$B_c=$$\begin{bmatrix}
       0            \\
			0            \\
       0             \\
       0             \\
			h_d^{-1}           
      \end{bmatrix}$
\end{centering}\\
\\
Since the lead vehicle does not follow any vehicles, it will not receive any information through wireless or its radar.
As a result, the lead vehicle dynamics will be different from the other vehicles' dynamics in the platoon.
The lead vehicle dynamics is defined by $x_0$ as
\begin{equation}
\begin{aligned}
\label{equ7}
\dot x_0 = A_0 x_0 + B_s u_l
\end{aligned}
\end{equation}
where

\quad \quad \quad \quad \quad \quad \quad \quad \quad \quad \quad \quad \quad \quad \quad \quad $A_0=$$\begin{bmatrix}
   0 &    0 & 0  &      0      &       0        \\
		0 &	0 & 0  &      0      &       0         \\
   0 &    0 &  0  &      1       &       0         \\
    0 &   0 &  0  & -\eta_0^{-1} &      0         \\
		0 &	 0 &  0  &      0       & 0        
     \end{bmatrix},$
	\quad	$B_s=$$\begin{bmatrix}
       0            \\
			0            \\
       0             \\
       -\eta_0^{-1}             \\
			0           
      \end{bmatrix}$

\subsection{Vehicles String State Space Representation}
The state space representation of the CACC control structure in a vehicle string is as follows \cite{SJ2014}:
\begin{equation}
\begin{aligned}
\label{equ8}
\dot{\bar{x}}_n =\bar{A}_n \bar{x}_n + \bar{B}_c \tilde u_{n-1} +\bar{B}_s u_{l}
\end{aligned}
\end{equation}	

where $u_{l}$ is an arbitrary commanded acceleration taken by the lead vehicle and

\quad \quad  $\bar{A}_n=$$\begin{bmatrix}
       A_0        &       0     &     0       &   \cdots     &   0        \\
			 A_{1,0}    &    A_{1,1}  &     0       &   \cdots   &   0        \\
       0          &    A_{2,1}  &   A_{2,2}   &   \cdots     &   0         \\
       \vdots  &                &   \ddots    &    \ddots  &  \vdots    \\
			 0          &   \cdots              &      0  &  A_{n,n-1} & A_{n,n}    \\       
     \end{bmatrix}_{5(n+1)\times 5(n+1)}$
$\bar {B}_{c}=$$\begin{bmatrix}
       0   &    0    &   \cdots   & 0        \\
			 0   &   B_c   &   \cdots   & 0        \\
       0   &    0    &   \ddots   & 0        \\
       0   &    0    &   \cdots   & B_c        \\
			         
     \end{bmatrix}_{5(n+1)\times(n+1)}$
		$\bar{B}_s=$$\begin{bmatrix}
       B_{s}           \\
			0            \\
       0             \\
       0             \\
			 0             \\
      \end{bmatrix}_{5(n+1) \times 1}$

and ${\bar{x}}_n=[x_0^T \quad x_1^T \quad x_2^T \quad . . . \quad x_n^T]^T$ represents the augmented state space variables of the vehicles' dynamics in the string.
In ~(\ref{equ8}), $\tilde u_{n-1}=[0 \quad \tilde u_{0} \quad \tilde u_{1} \quad . . . \quad \tilde u_{n-1}]^T$ is a vector where its elements denote the received acceleration information of $ith$ vehicle (for $i=0,...,n-1$) in its immediately following vehicle.
The first element in the vector (zero value) indicates that the lead vehicle does not receive any acceleration information. 

Considering that the DSRC transmission policy is based on sending out the data every $100ms$ over the wireless network \cite{Li2012,SN2014}, the signal $\tilde{u}_{i-1}$ at times $t_k=kh$ is sampled for $k=0,1,2,...$ and $h=100 ms$, to represent the DSRC functionality.
This also means that if the packets are received successfully, then the following vehicle's receiver will get updated in a fixed periodic transmission manner. 
The following state space representation captures the signal sampling and holding it by the ZOH in the receiver. 
 
\begin{equation}
\begin{aligned} 
\label{equ9}
\dot{\bar{x}}_n =\bar{A}_n \bar{x}_n + \bar{B}_c \tilde u_{n-1} +\bar{B}_s u_{l}\\
\tilde u_{n-1}(t) = \tilde u_{n-1,k}, \quad t\in[t_k, t_{k+1}]
\end{aligned}
\end{equation}
where $\tilde u_{n-1,k} = \tilde u_{n}(t_k)$. 
Now we compute the exact discrete-time representation for the continuous-time system in~(\ref{equ9}) as follows:
\begin{equation}
\begin{aligned}
\label{equ10}
\bar{\bar{x}}_n[k+1] = \bar{\bar{A}}_n \bar{\bar{x}}_n[k]+\bar{\bar{B}}_c \tilde u_{n-1}[k] + \bar{\bar{B}}_s u_{l}[k]\\
\bar{\bar{A}}_n= e^{\bar{A}_nh},\quad \bar{\bar{B}}_c = \int_{0}^{h}e^{\bar{A}_n\nu} d\nu . \bar{B}_c,\quad \bar{\bar{B}}_s= \int_{0}^{h}e^{\bar{A}_n\nu} d\nu . \bar{B}_s  \\
\end{aligned}
\end{equation}
where $h$ is the sampling interval. $\bar{\bar{A}}_n$, $\bar{\bar{B}}_c$ and $\bar{\bar{B}}_s$ are the time-invariant matrices computed for the discreet-time string state space representation in ~(\ref{equ10}).

\subsection{String Stability} 
The lead vehicle's acceleration and deceleration will produce spacing error $e_i$ between $i$th and ($i-1$)th vehicle in the platoon for $i=1,2,...,n$.
String stability requires spacing error attenuation along the vehicle string. 
In other words, a string will be stable if the generated spacing error as a result of the lead vehicle's action does not get amplified when it propagates upstream the string. 
This requirement is expressed as follows \cite{XA2001}:
\begin{equation}
\label{equ11}
\lVert e_n \rVert_\infty< \lVert e_{n-1} \rVert_\infty <...< \lVert e_2 \rVert_\infty < \lVert e_1 \rVert_\infty 
\end{equation}
where $\lVert . \rVert_\infty$ denotes the infinity norm which determines the maximum absolute value of the corresponding spacing error in a time horizon of $t$. In other words, the time-domain definition of the string stability is given by

\begin{equation}
\label{equ12}
\begin{aligned}
\text{max}_{t}\left|e_n(t)\right|&< \text{max}_{t}\left|e_{n-1}(t)\right| < ... \\
&...< \text{max}_{t}\left|e_2(t)\right| < \text{max}_{t}\left|e_1(t)\right|\\
\end{aligned}
\end{equation}

Equation~(\ref{equ12}) indicates that string will be stable if and only if the maximum absolute value of the produced error in a time horizon of $t$ gets diminished as it propagates upstream the string.

When the dynamic of the CACC system is deterministic as in (\ref{equ9}), string stability is evaluated by the frequency domain definition and string will be stable if the following condition is satisfied \cite{GR2010}:
\begin{equation} 
\label{equ13}
\left| \Gamma(j\omega) \right| = \left|\frac{E_i(j\omega)}{E_{i-1}(j\omega)}\right|\leq 1 \quad \forall \omega, \quad i=1,...,n
\end{equation}
where $E_i(j\omega)=\mathcal{F}(e_i(t))$ represents the Fourier transformation of the spacing error for the $i$th vehicle. 

However, when communication uncertainty is introduced into the CACC dynamics, state space variables will be stochastic variables which determine the CACC system's behavior. 
For clarity, we use bold letters to denote the stochastic state space variables.  
In order to study the string stability while considering stochastic dynamics for the CACC systems, the concept of mean string stability is defined as
\begin{equation}
\label{equ14}
\begin{aligned}
\mathbb{E}\left\{\text{max}_{t}\left|\bold{e_n}(t)\right|\right\}&< \mathbb{E}\left\{\text{max}_{t}\left|\bold{e_{n-1}}(t)\right| \right\}< ... \\
&...< \mathbb{E}\left\{\text{max}_{t}\left|\bold{e_2}(t)\right|\right\} < \mathbb{E}\left\{\text{max}_{t}\left|\bold{e_1}(t)\right|\right\}\\
\end{aligned}
\end{equation}
where $\mathbb{E}\left\{. \right\}$ represents the expected value.

In the next section, we will incorporate jamming attack and wireless channel condition effects into ~(\ref{equ10}) in order to analyze the mean string stability and reachable inter-vehicle distance states for safety verification.    
\section{Integrating Jamming Attack and Rician Fading into CACC Model}
\label{AttackFading}       
In this section, we model jamming attack and Rician fading impact on the state space representation of the string.    
The model captures the dependency of the physical states (inter-vehicle distances) on the cyber part (unreliable wireless channel states) and vice versa. 
Each packet sent by the vehicles can be lost due to Rician fading or the attacker's destructive signal. 
This uncertain packet delivery affects the error propagation along the vehicle string and the inter-vehicle distance states evolution in the platoon.

\subsection{Attack Model}
Basically practical studying of behavior of vehicular platooning requires considering real scenarios.
Because of the vehicles' mobility, the jammer is considered as a mobile attacker.       
Recall that we assume a reactive jammer  mounted on a drone flying over the platoon emits its jamming signal over the wireless network whenever it senses that the communication traffic is happening in the network \cite{WW2005}.
The jammer's destructive signal is considered as an additive Gaussian random variable $J \sim \mathcal{N}(\mu_j,\,\sigma_j^{2})\,$ with known and constant mean ($\mu_j$) and variance ($\sigma_j^{2}$). 
The jammer's transmitted signal power is calculated as $P_j = \left|\mu_j\right|^2 + \sigma_j^{2}$, \cite{LW2016}. 
This jamming model is a flexible model for representing a wide range of jamming signal scenarios.   
The ratio $M = \frac{\left|\mu_j\right|^2}{\sigma_j^{2}}$ represents the jamming signal's features in terms of signal's power.
For example, when $M=0$ or $\mu_j = 0$, the jamming signal becomes a zero-mean Gaussian random variable which generates a very powerless noise signal rather than a strong jamming signal.
Whereas, when $M \rightarrow \infty$ or $\sigma_j^{2}=0$, jamming signal appears as a constant jamming signal with Additive White Gaussian Noise (AWGN) in the model.  
However, the general scenario will be the case that $0 \leq M \leq \infty$. 
In this case, there is a strong jamming signal with noise in the medium that jammer's antenna beam covers.   

The mean power of the jammer's signal at the receiver of the $i$th vehicle at time $k$ considering free space path loss model is obtained as 
\begin{align}
\label{equ15}
I_{i}^k = \frac{G_j G_r \lambda^2(\left|\mu_j\right|^2 + \sigma_j^{2})}{(4 \pi)^2 (s_{i}^k)^\alpha}
\end{align}
where, 
\begin{align}
\label{equ16}
s_{i}^k =\left \{ \begin{array}{cc}
\sqrt{(\sum_{m=i+1}^{j}d_m^k)^2+l^2},    &   i \leq j-1    \\ 
l,               &  i = j  \\
\sqrt{(\sum_{m=j+1}^{i}d_m^k)^2+l^2},       &i \geq j+1
\end{array}  \right. 
\end{align}
for $i,j=1,2,...,n$. 

$G_j$ and $G_r$ denote the jammer and vehicles' receiver antenna gain, respectively. 
$\alpha$ indicates the path loss exponent, $\lambda = c/f_0$ is the associated wavelength ($c$ is speed of light and $f_0$ is the carrier frequency).
$l$ denotes the drone's vertical distance from the platoon (Figure~\ref{platoon}).
In~(\ref{equ16}), $s_{i}^k$ represents jammer distance from $i$th vehicle when the jammer is located above the $j$th vehicle in the platoon at time $k$.

According to the velocity-dependent spacing policy, if the lead vehicle's velocity increases, the actual distance between a pair of vehicles in the platoon increases as well and vice versa. 
Therefore, the distance between the jammer and each vehicle in the platoon ($s_{i}^k$) depends on the lead vehicle's acceleration profile. 
Consequently, we conclude that since $I_{i}^k$ is a function of $s_{i}^k$, then the mean power of the jammer's signal at the receiver of each vehicle is also influenced by the lead vehicle's acceleration profile.
 
\subsection{Attack and Rician Fading Model Integration}   
As the distance between each pair of transmitter and receiver keeps changing, received signal strength is also varying. 
We consider free space path loss model for the received signal's power level variation with respect to the instant inter-vehicle distance. 
Thus, the received signal power $P_{r,i}^k$ at the receiver of the $i$th vehicle at time $k$ is expressed as 
\begin{align}
\label{equ17}
P_{r,i}^k = \frac{G_t G_r \lambda^2 P_{t,i}^k} {(4 \pi)^2 (d_i^k)^\alpha} 
\end{align}
where $P_{t,i}^k$ and $G_t$ denote the transmission signal power of the $i$th vehicle and transmitter antenna gain, respectively.

We consider the jammer's signal as an interference signal that is added to the ambient noise.
Thus, we compute signal-to-interference-plus-noise ratio (SINR) to derive the probability of successful packet delivery.
Once the attacker launches its jamming signal over the platoon, average SINR of the received signal of the $i$th vehicle at time $k$ is derived by

\begin{align}
\label{equ18} 
\overline{\gamma_i^k}=\overline{SINR_i^k} = \frac{P_{r,i}^k}{(N_0+I_{i}^k)} = \frac{G_t G_r \lambda^2 P_{t,i}^k} {(4 \pi)^2 (d_i^k)^\alpha(\sigma_n^2 + \frac{G_j G_r \lambda^2(\left|\mu_j\right|^2 + \sigma_j^{2})}{(4 \pi)^2 (s_{i}^k)^\alpha})}
\end{align}
where $N_0 = \sigma_n^2$ represents the mean power of ambient noise which is considered as an additive Gaussian random variable with zero mean and variance $\sigma_n^2$. 
From (\ref{equ18}), we see that the average SINR is a function of distance states ($d_i^k , s_{i}^k$), which indeed indicates the dependency of wireless channel quality on the system states. 

As stated in the system model, each vehicle in the platoon receives its immediate preceding vehicle's commanded acceleration information.
Thus, due to the possessing line-of-sight communication channel between each transmitter and receiver, Rician fading channel model is considered as a fairly good stochastic model for this class of signal transmission and environment. 
In this model, received signal amplitude has a Rician probability density function.
With the assumption of Rician fading, the probability density function of instantaneous SINR, $\gamma_i^k$, is expressed as follows \cite{AB1994, WL2009}:

\begin{align}
\label{equ19}
f_{i}^k(\gamma_i^k) = \frac{1+K}{\overline{\gamma_i^k}}exp\left(-K-\frac{\left(1+K\right)\gamma_i^k}{\overline{\gamma_i^k}}\right) \times I_{0}\left(2\sqrt{\frac{K(K+1)\gamma_i^k}{\overline{\gamma_i^k}}}\right)
\end{align}
where $K$ is the Rician fading parameter and represents the ratio of the received signal's power in the LOS component to the non-LOS scattered multipath components. When $K\longrightarrow \infty$, the channel is equivalent to a static additive white Gaussian noise (AWGN) channel, and with $K = 0$, the channel reduces to Rayleigh fading channel.
$I_0(.)$ is the zero order modified Bassel function of the first kind.

To decode a received packet successfully, the instantaneous SINR should be greater than an acceptable SINR ($\gamma_{th}$) \cite{AB1994}. 
Therefore, the probability of successful packet delivery is defined as follows:

\begin{align}
\label{equ20}
 p_{i-1}^k = \bold{P}_{i-1}^k(\gamma_i^k \geq \gamma_{th})= 1- F_{\gamma_i^k}(\gamma_{th}) = Q \left(\sqrt{2K},\sqrt{\frac{2(1+K)\gamma_{th}}{\overline{\gamma_i^k}}}\right)\quad \text{for} \quad i=1,...,n
\end{align}
where $\bold{P}$ denotes the probability. $Q(.,.)$ and $F_{\gamma_i^k}(\gamma_{th})$ are the first-order Marcum Q function and the cumulative distribution function (CDF) of the instantaneous SINR, respectively.
In ~(\ref{equ20}), $p_{i-1}^k$, for $i=1,...,n$, denotes the probability of successful packet delivery of the $(i-1)$th vehicle at the $i$th vehicle's receiver at time $k$.
In fact, $p_{i-1}^k$ has the opposite meaning of outage probability which is defined as the probability that the instantaneous SINR ($\gamma_i^k$) drops below the acceptable SINR ($\gamma_{th}$).
This probability is time variable and at each time, it depends on the average SINR derived in (\ref{equ18}), which is also an state dependent function. 
Other parameters such as transmitters' power, attacker's power and its distance from each receiver also affect the successful packet delivery probability.


Now, we define a Bernoulli random variable $\beta_{i-1}^k$ to indicate the packet successful delivery as follows:
\begin{align}
\label{equ21}
\beta_{i-1}^k=\left \{ \begin{array}{cc}
1,    &  p_{i-1}^k    \\ 
0,               &  1-p_{i-1}^k 
\end{array}  \right. 
\end{align}
for $k=1,2,...$ and $i=1,2,...,n$. 

Considering that each receiver has a memory unit (memory unit keeps the last successfully decoded acceleration information received from the immediate preceding vehicle and feeds it to the ZOH), $\tilde{u}_{i-1}[k]$ and its values backward in time are defined as follows: 

\begin{equation}
\begin{aligned} 
\label{equ22}
\tilde{u}_{i-1}[k] &= \beta_{i-1}^k u_{i-1}[k] + (1- \beta_{i-1}^k) \tilde{u}_{i-1}[k-1]\\
\tilde{u}_{i-1}[k-1] &= \beta_{i-1}^{k-1} u_{i-1}[k-1] + (1- \beta_{i-1}^{k-1}) \tilde{u}_{i-1}[k-2]\\
\vdots \\
\tilde{u}_{i-1}[2] &= \beta_{i-1}^2 u_{i-1}[2] + (1- \beta_{i-1}^2) \tilde{u}_{i-1}[1]\\
\tilde{u}_{i-1}[1] &= \beta_{i-1}^1 u_{i-1}[1] + (1- \beta_{i-1}^1) \tilde{u}_{i-1}[0]\\
\tilde{u}_{i-1}[0] &= \beta_{i-1}^0 u_{i-1}[0] \\
\end{aligned}
\end{equation}
for $i=1,2,...,n$ and $\beta_{i-1}^0 = 1$. 
Note that in (\ref{equ22}), ${u}_{i-1}[k]$ indicates the acceleration information of the ($i-1$)th vehicle before transmission whereas, $\tilde{u}_{i-1}[k]$ denotes the output of the memory unit in the CACC control structure.  
Considering the recursive format of ~(\ref{equ22}), $\tilde u_{i-1}[k]$ can be expressed as \\
\begin{equation}
\begin{aligned} 
\label{equ23}
\tilde{u}_{i-1}[k] = \beta_{i-1}^k u_{i-1}[k] + \sum_{m = 1}^{k} \beta_{i-1}^{m-1} u_{i-1}[m-1] \prod_{j=m}^{k}(1-\beta_{i-1}^{j}) 
\end{aligned}
\end{equation}
Therefore, state space representation of the platoon under Rician fading channel and jamming attacks is derived as follows:

\begin{equation}
\begin{aligned}
\label{equ24}
\bold{x_n}[k+1] = \bar{\bar{A}}_n \bold{x_n}[k]+\bar{\bar{B}}_c \tilde u_{n-1}[k] + \bar{\bar{B}}_s u_{l}[k]\\
\end{aligned}
\end{equation}
where

\quad \quad \quad \quad \quad \quad \quad \quad \quad $\tilde u_{n-1}[k]=\begin{bmatrix} 
 
          0                                                                                                          \\
       \beta_0^k u_l[k] + \sum_{m = 1}^{k} \beta_0^{m-1} u_l[m-1] \prod_{j=m}^{k}(1-\beta_{0}^{j})               \\
			 \beta_1^k u_1[k] + \sum_{m = 1}^{k} \beta_1^{m-1} u_1[m-1] \prod_{j=m}^{k}(1-\beta_{1}^{j})           \\
      \vdots           \\
			 \beta_{n-1}^k u_{n-1}[k] + \sum_{m = 1}^{k} \beta_{n-1}^{m-1} u_{n-1}[m-1] \prod_{j=m}^{k}(1-\beta_{n-1}^{j})               \\         
			
     \end{bmatrix}_{(n+1)\times 1}$\\\\
and $u_{l}$ is an arbitrary commanded acceleration taken by the lead vehicle.

Equation~(\ref{equ24}) shows that except for the lead vehicle commanded acceleration $u_l$, other vehicles commanded acceleration value will be a random variable and are computed recursively with respect to time. 

We will use the stochastic dynamical system of the CACC system presented in (\ref{equ24}) to study the wireless communication uncertainty and jamming attacks impact on the string stability and safety.  
\section{String Stability Analysis} 
\label{StringStabilityAnalysis}      
Existing string stability analysis are based on frequency-domain techniques \cite{GR2010,YK1996,QG2017,EJ2007}.
However, for the stochastic dynamical system obtained in (\ref{equ24}), frequency-domain analysis method cannot be employed because of time-varying probabilistic packet successful delivery at each receiver.
In order to tackle this challenge, we employ time-domain method for analysis of string stability and distance states evolution of CACC system under Rician fading channel and jamming attacks. 
Next, first we validate the time-domain method through comparing it with frequency-domain method for the case of perfect channel condition (No fading, No attack), and normal operation of the platoon.
Then, for the remained subsections, time-domain analysis is utilized to evaluate the impact of Rician fading channel and jamming attacks on the CACC performance and functionality.
\begin{table}%
\caption{Simulation Configuration}
\squeezeup
\label{tabSim1}
\begin{minipage}{\columnwidth}
\begin{center}
\begin{tabular}{ll}
  \toprule 
  Number of vehicles following the lead vehicle              & $n = 10$\\
  Internal actuator dynamics                     & $\eta_i=\eta = 0.1$\\
  Derivative constant                       & $k_d=0.5$ \\
 Proportional constant                      & $k_p=0.25$\\
			Carrier frequency                           & $f_0 = 5.9$ [GHz]\\
\bottomrule
\end{tabular}
\squeezeup
\end{center}
\bigskip\centering
\end{minipage}
\end{table}%

\subsection {CACC System Settings}
We consider a platoon constructed with $n=10$ vehicles plus the lead vehicle. 
The lead vehicle's index is zero and the rest of the vehicles are ordered from one to ten moving down the platoon.   
We assume that the vehicles are homogenous and the internal actuator dynamics are identical for all vehicles in the platoon ($\eta_i=\eta=0.1$ for $i=0,1,2,...,n$).
Also, $k_{d,i}=k_d=0.5<<1/\eta$ and $k_{p,i}=k_p=k_d^2=0.25$ for $i=1,...,n$ are chosen to satisfy the internal stability of the vehicle dynamics. 
A summary of simulation configuration are listed in Table \ref{tabSim1}. 
The simulation set up parameters' value mentioned in Table \ref{tabSim1} are fixed through out all the remained sections. 
Simulation computations are performed with Matlab software.

\subsection {String Stability Analysis and Headway-time Optimization (No Fading and No Attack)}

We assume perfect channel condition, no  fading and no attack scenarios, for all the V2V wireless communication channels in the platoon.
In other words, all the packets are delivered and decoded at each receiver without being delayed or dropped.  
To analyze the string stability in the frequency domain, we derive transfer function of string stability for the CACC and ACC (CACC without V2V communication) modes using the control structure shown in Figure~\ref{CACCmodelattack}.    
By solving the following non-linear deterministic optimization problem, the minimum headway-time is obtained by which string stability is guaranteed.

\begin{figure}[!htb]
    \centering
    \begin{minipage}{.5\textwidth}
        \centering
        \includegraphics[width=1\linewidth, height=0.3\textheight]{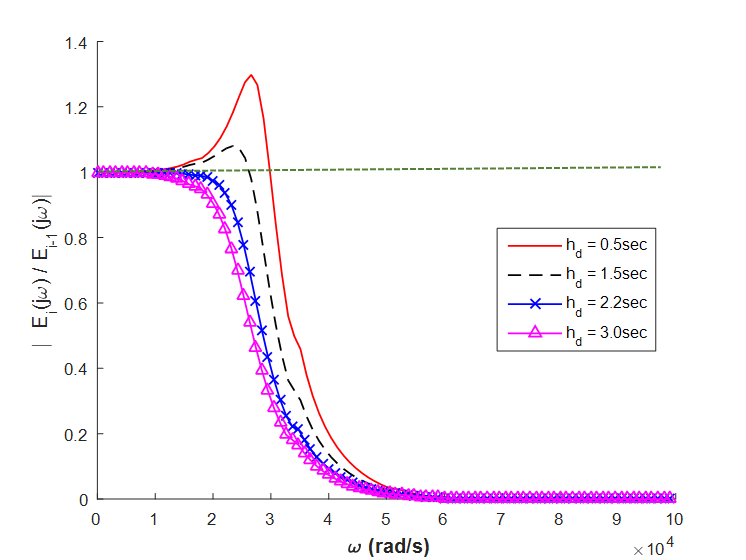}
        \subcaption{ Frequency-domain string stability for ACC mode}
        \label{fig:prob1_6_2}
    \end{minipage}%
    \begin{minipage}{0.5\textwidth}
        \centering
        \includegraphics[width=1\linewidth, height=0.3\textheight]{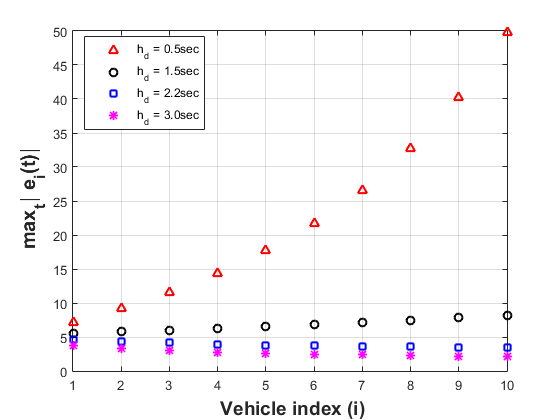}
        \subcaption{Time-domain string stability for ACC mode}
        \label{fig:prob1_6_1}
    \end{minipage}
		\\
		 \begin{minipage}{.5\textwidth}
        \centering
        \includegraphics[width=1\linewidth, height=0.3\textheight]{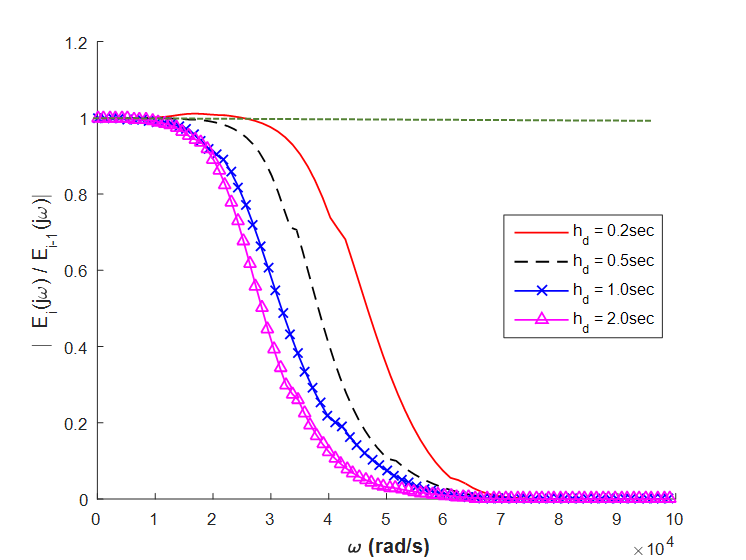}
        \subcaption{Frequency-domain string stability for CACC mode}
        \label{fig:prob1_6_2}
    \end{minipage}%
    \begin{minipage}{0.5\textwidth}
        \centering
        \includegraphics[width=1\linewidth, height=0.3\textheight]{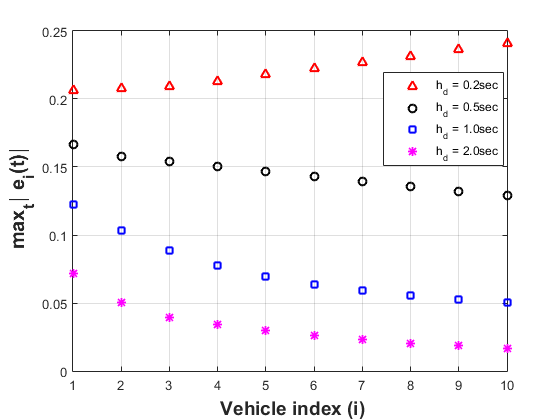}
        \subcaption{Time-domain string stability for CACC mode}
        \label{fig:prob1_6_1}
    \end{minipage}
		
\caption{Frequency and Time domain stability for ACC and CACC}
\squeezeup
\label{Simluton}
\end{figure}
 
\begin{figure}[!htb]
    \centering
    \begin{minipage}{.5\textwidth}
        \centering
        \includegraphics[width=1\linewidth, height=0.3\textheight]{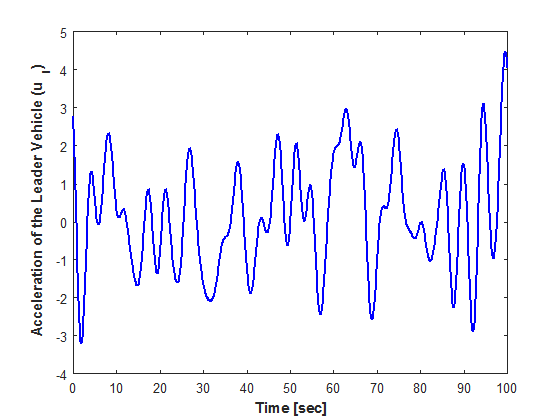}
        \subcaption{Sample lead vehicle acceleration profile}
        \label{fig:prob1_6_2}
    \end{minipage}%
    \begin{minipage}{0.5\textwidth}
        \centering
        \includegraphics[width=1\linewidth, height=0.3\textheight]{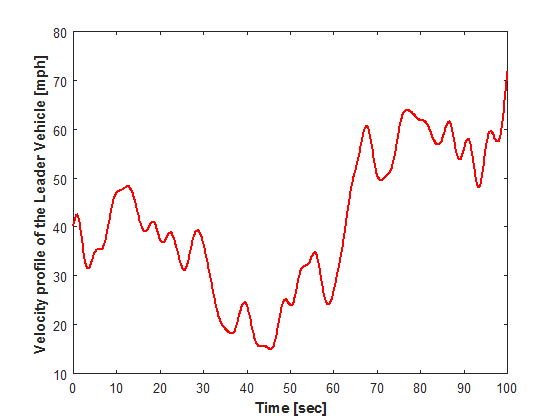}
        \subcaption{ Sample lead vehicle velocity profile}
        \label{fig:prob1_6_1}
    \end{minipage}
		\caption{}
\squeezeup
\label{profile}
\end{figure}

\begin{equation}
\begin{aligned}
\label{equ25}  
&{\text{minimize}} 
&&  h_d \nn
& \text{subject to} 
&&  \omega \geq 0 \nn
&&& h_d > 0 \nn
&&&  \left|\frac{E_i(j\omega)}{E_{i-1}(j\omega)}\right|\leq 1 \quad \text{for} \quad i=1,2,...n
\end{aligned}
\end{equation}
where,
\begin{equation}
\begin{aligned}
\label{equ26}
\frac{E_i(j\omega)}{E_{i-1}(j\omega)} = \frac{k_{d,i}j \omega + k_{p,i}}{\eta_i (j \omega)^3 + (k_{d,i}+1)(j \omega)^2 + (k_{d,i} + k_{p,i} h)(j \omega) + k_{p,i}} 
\end{aligned}
\end{equation}
and,
\begin{equation}
\begin{aligned}
\label{equ27}
\frac{E_i(j\omega)}{E_{i-1}(j\omega)} = \frac{\eta_i (j \omega)^3 + (1+k_{d,i})(j \omega)^2 + (k_{d,i} + k{p,i} h)(j \omega) + k_{p,i}}{\eta_i h_d (j \omega)^4 + (k_{d,i} h_d^2 + h_d + \eta_i)(j \omega)^3 + (k_{p,i} h_d ^ 2 + h_d + 1)(j \omega)^2 + (k_{d,i} h_d + k_{d,i})(j \omega) + k_{p,i}}  
\end{aligned}
\end{equation} 
are string stability transfer functions for the ACC and CACC modes, respectively.
Optimization problem is solved via GAMS software \cite{GAMS2013}.
Minimum headway-time $h_d$ for the ACC and CACC modes are obtained as $2.101$ seconds and $0.284$ seconds, respectively.  
Figures~\ref{Simluton}(a) and~\ref{Simluton}(c) illustrate the string stability analysis of ACC and CACC systems in the frequency domain.
These results, show the string stability transfer function's absolute magnitude for various headway-times against the wide range of frequencies ($0-10^5 rad/s$).
Using the string stability definition in ~(\ref{equ13}), in Figures~\ref{Simluton}(a) and~\ref{Simluton}(c), for the headway times for which the absolute value of $\frac{E_i(j\omega)}{E_{i-1}(j\omega)}$ exceeds the value of $1$, string becomes unstable.
To demonstrate the performance of ACC and CACC systems under equal settings, a video of full demo is displayed at \url{https://youtu.be/B1ls0HaGULs}.

Now, we analyze the string stability in the time-domain and validate the results by comparing them with the results of the frequency domain analysis.
We generate commanded acceleration profiles for the lead vehicle using the random phase multi-sine signal generation method \cite{FJ2004, PS2005}.
This method has been used in \cite{SJ2014} to generate velocity profile for the lead vehicle. 
Produced acceleration profiles model the lead vehicle's real-world actions. 
One sample of the lead vehicle's acceleration and the corresponding velocity profile up to $100$ seconds are shown in Figures~\ref{profile}(a) and~\ref{profile}(b), respectively.  
		
In the time-domain, maximum spacing error produced at each vehicle in a time horizon of $500$ seconds is computed for the ACC and CACC modes for various headway-times. 
The results are illustrated in Figures~\ref{Simluton}(b) and~\ref{Simluton}(d).   

For the case of ACC system, by comparing the results in both frequency and time domains, Figures~\ref{Simluton}(a) and~\ref{Simluton}(b), we observe that when the headway-time is below 1.5 seconds the string gets unstable.
However, string is stable for both domains when the headway-time is set to 2.2 seconds and 3 seconds. 
For the case of CACC system, Figures~\ref{Simluton}(c) and~\ref{Simluton}(d) show that in both domains for the headway-time of 0.2 seconds, string is unstable. However, string is stable for the headway times of 0.5 seconds, 1 second and 2 seconds.
The results are consistent with the solutions obtained from the optimization problem in (\ref{equ25}).
Through on the comparison of the analysis, we conclude that string stability analysis of both frequency-domain and time-domain are highly consistent and endorse each other.
Based on the time and frequency domains consistency, next we aim to analyze the mean string stability and reachable inter-vehicle distance states in the time-domain when V2V wireless communication is subject to Rician fading channel and jamming attacks.
 
\begin{table}%
\caption{Simulation Configuration}
\squeezeup
\label{tabSim2}
\begin{minipage}{\columnwidth}
\begin{center}
\begin{tabular}{ll}
  \toprule 
  Vehicles initial velocity                       & $v_i[0] = 40 $ [mph]  for $i=0,...,n$\\   
  Vehicles initial commanded acceleration        & $2.70  [m/s^2]$\\
  Vehicle initial spacing error                  & $e_i[0] = 0 $ \\   
  Headway time                                   & 1 second\\
  Jamming signal mean                     & $\mu_j = 0.00173$\\
	Jamming signal variance                     & $\sigma_{j} = 0.001$ \\
	Noise power                              & $N_0 = -80 $ [dBm]\\
	Constant Vehicle's transmission power     & $P_{t,i}= 28 $ [dbm]for $i=0,...,n$\\
  Threshold SINR                           &  $\gamma_{th} = 18 $dB\\
	Transmitter and aeceiver antenna gains   & $G_t=G_r= 12 $ [dBi]\\
	Jammer anntenna gain                     & $G_j = 18$ [dBi] \\
	path loss component                       & $\alpha = 2$\\
	Vertical distance of drone from platoon      & $l=6 $ m\\
	Rician fading parameter                          & $K = 2$\\
\bottomrule
\end{tabular}
\end{center}
\bigskip\centering
\end{minipage}
\end{table}%

\subsection {Impact of Jamming Attack } 
In this subsection, we study the impact of jamming signal power and the attacker's location on the string stability. 
We assume that the jammer is mounted on a drone, flying over the platoon.
The attacker emits its jamming signal over the platoon wireless communication network in the whole time horizon.   
We assume that the attacker (i.e. drone) has been equipped with an appropriate adaptive velocity controller such that it has the capability of maintaining the same speed as the vehicles in the platoon. 
The drone also maintains a fixed altitude from the platoon. 
We also assume that the signal transmission power for all the vehicles ($i=0,1,...,10$) is fixed ($P_{t,i} = 28 [dBm]$) and identical on all the time for all the scenarios considered in this subsection.
The mobile jammer's signal power is also fixed, $P_j = -24 [dBm]$, all the time for all scenarios in this subsection.
We also set the head-way time to $1$ second ($h_d=1$ second) in the control structure of CACC system.
Table \ref{tabSim2}, summarizes the simulation parameters' value complementary to the Table \ref{tabSim1}.

We generate $1,000$ acceleration profiles to cover as many as possible actions that the lead vehicle can take.
The profiles are generated using the random phase multi-sine signal generation method \cite{FJ2004, PS2005} with profile duration of $500$ seconds.
For each acceleration profile, we run the simulations $10,000$ times for each scenario.
Finally, the mean maximum spacing error for the $i$th vehicle is computed as follows: 

\begin{equation}
\label{equ101} 
\begin{aligned}
\mathbb{E}\left\{\text{max}_{t}\left|\bold{e_i}(t)\right|\right\} = \frac{1}{N_p} \sum_{p=1}^{N_p} \left(\frac{1}{N_q} \sum_{q=1}^{N_q} \left(\text{max}_{t}|e_{i_{pq}}(t)|\right)\right)  \quad \text{for} \quad i = 1,2,...n
\end{aligned}
\end{equation}
where $N_p$ and $N_q$ are the number of commanded acceleration profiles and iterations, respectively. 
$e_{i_{pq}}(t)$ denotes the spacing error generated for the $p$th profile at iteration $q$ for the $i$th vehicle  in a time horizon of $t$ seconds.
After computing $\mathbb{E}\left\{\text{max}_{t}\left|\bold{e_i}(t)\right|\right\}$ \text{for} $i=1,2,...,n$, we use (\ref{equ14}) to study the mean string stability. 

\begin{figure}[!htb]
    \centering
    \begin{minipage}{.5\textwidth}
        \centering
        \includegraphics[width=1\linewidth, height=0.3\textheight]{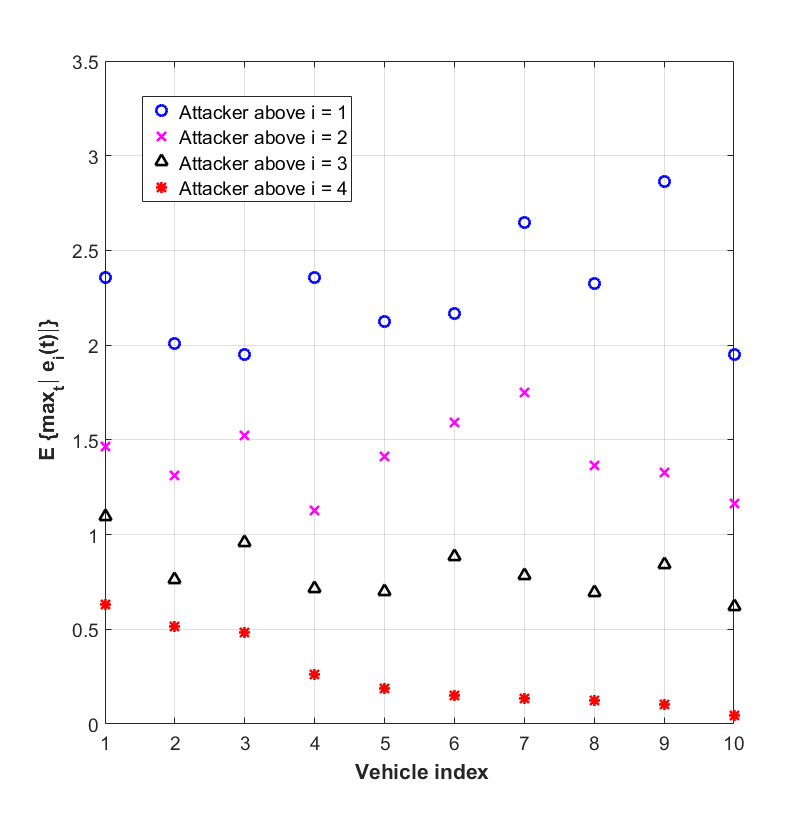}
        \subcaption{Attacker location impact on mean string stability}
        \label{fig:prob1_6_2}
    \end{minipage}%
    \begin{minipage}{0.5\textwidth}
        \centering
        \includegraphics[width=1\linewidth, height=0.3\textheight]{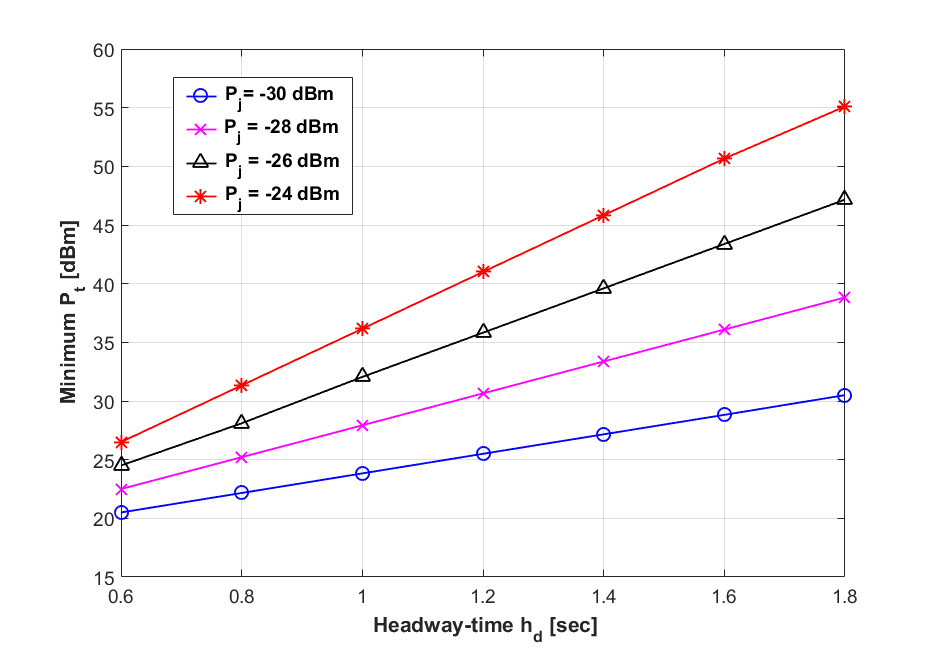}
        \subcaption{Minimum transmission power vs. headway-time}
        \label{fig:prob1_6_1}
    \end{minipage}
		\caption{Mean String Stability and Minimum Transmission Power}
\squeezeup
\label{attackpower} 
\end{figure}

Figure~\ref{attackpower}(a) demonstrates the jammer's capability of destabilizing the platoon. 
As the results show in Figure~\ref{attackpower}(a), when the attacker is above the second vehicle ($i=1$), not only the mean maximum error oscillates upstream the string, but also the magnitude of the errors are larger in comparison to the no attack scenarios shown in Figure~\ref{Simluton}(d).
The results in Figure~\ref{attackpower}(a) show that, as the attacker moves upstream in the platoon, its ability to destabilize the platoon diminishes. 
This is because as the attacker moves far away from the lead vehicle, produced spacing error at the front vehicles are decreased since the packet delivery ratio increases.
As a result, the more the attacker moves away from the lead vehicle, the more spacing error is corrected by the CACC controllers such that when the attacker is above the forth vehicle in the platoon, the string becomes stable. 
Therefore, we conclude that the closer the attacker gets to the second vehicle  ($i=1$ or first vehicle following the lead vehicle), the more effective it will be in terms of destabilizing the platoon.

\subsection {Minimum Transmission Power to Maintain Mean String Stability}
 
After finding the best location (above $i=1$) for the attacker in the previous part, in this subsection, we aim to study the jammer's and the vehicles' transmission powers impacted by the headway-time variation on the mean string stability. 
Therefore, as a defending strategy against the jamming signal, we find the vehicles minimum required transmission power such that for a jammer with fixed power and location (above $i=1$), the platoon maintains the mean stability.
Minimum transmission power is obtained as follows:
 
\begin{equation}
\begin{aligned}
\label{equ102} 
&{\text{minimize}} \quad \quad \quad P_t \\
&\text{subject to} \quad \quad \quad \mathbb{E}\left\{\text{max}_{t}\left|\bold{e_n}(t)\right|\right\}< \mathbb{E}\left\{\text{max}_{t}\left|\bold{e_{n-1}}(t)\right| \right\}< ... \\
&\quad \quad \quad \quad \quad \quad \quad \quad \quad \quad \quad \quad \quad ...< \mathbb{E}\left\{\text{max}_{t}\left|\bold{e_2}(t)\right|\right\} < \mathbb{E}\left\{\text{max}_{t}\left|\bold{e_1}(t)\right|\right\} 
\end{aligned}
\end{equation} 
This problem is solved by simulation under the same system setting parameters used in mean string stability analysis. 
The results in Figure~\ref{attackpower}(b) illustrates that for a fixed headway-time, as the jamming signal power increases, the vehicles minimum transmission power need to be increased as well to stabilize the string.
In addition, assuming a fixed signal power for the jammer, as the headway-time increases, a higher transmission power is required for the vehicles to maintain the mean string stability.
This is because, based on the velocity dependent spacing policy, with a bigger headway-time, vehicles are moving with larger inter-vehicle distance compared to the one with smaller headway-time.

\begin{figure}[!htb]
    \centering
    \begin{minipage}{.5\textwidth} 
        \centering
        \includegraphics[width=1\linewidth, height=0.3\textheight]{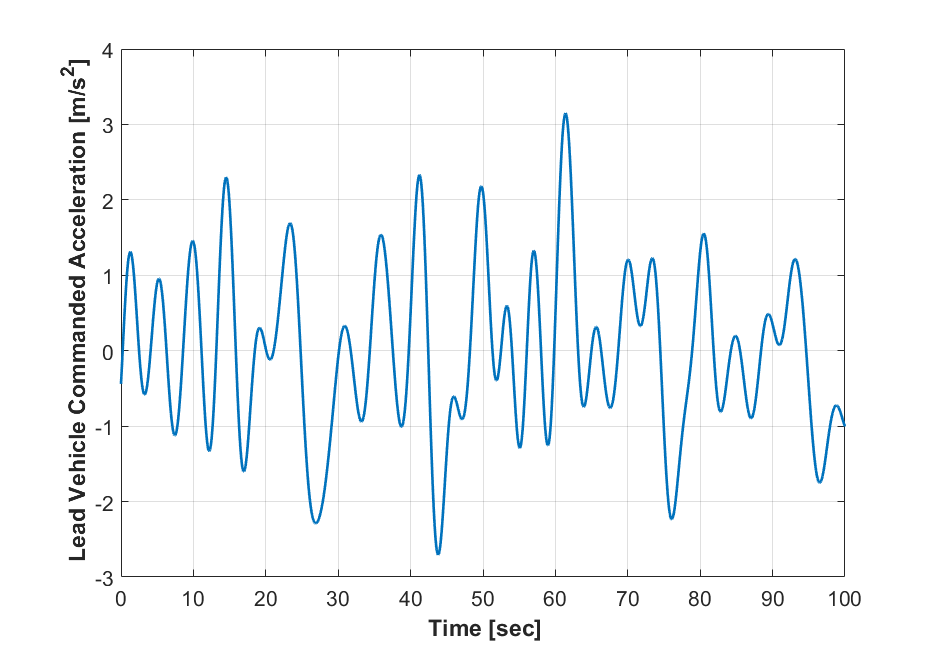}
        \caption{Lead vehicle commanded acceleration profile}
        \label{lead}
    \end{minipage}
\squeezeup 
\end{figure}   

\section{Safety Evaluation Under Channel Faiding and Jamming Attacks}
\label{Distance Reachable Sets}   

In this section, to examine the reachable inter-vehicle distance states at each time instant, we define a vector $\bold{y}[k] = G \bold{x_n}[k]$, where $\bold{y}[k]=[\bold{d_0}[k] \quad \bold{d_1}[k] \quad \bold{d_2}[k] \quad ... \quad \bold{d_n-1}[k] \quad \bold{d_n}[k]]$ and $\bold{x_n}[k]$ is obtained from (\ref{equ24}). 
$G$ is a matrix with dimension of $(n+1)\times 5(n+1)$ with all elements equal to zero except the elements in $(i,5i-4)$ for $i =1,2,...n+1$, which are equal to 1.   
 
\subsection {Impact of Rician Fading Channel on Inter-Vehicle Distances}  
In this subsection, we study the reachable inter-vehicle distance states in the platoon when V2V wireless communication channels suffer from Rician fading.
We assume that there is no attacker in the system model. 
\begin{figure}[!htb]
    \centering
    \begin{minipage}{.31\textwidth}
        \centering
				\includegraphics[width=1\linewidth, height=0.17\textheight]{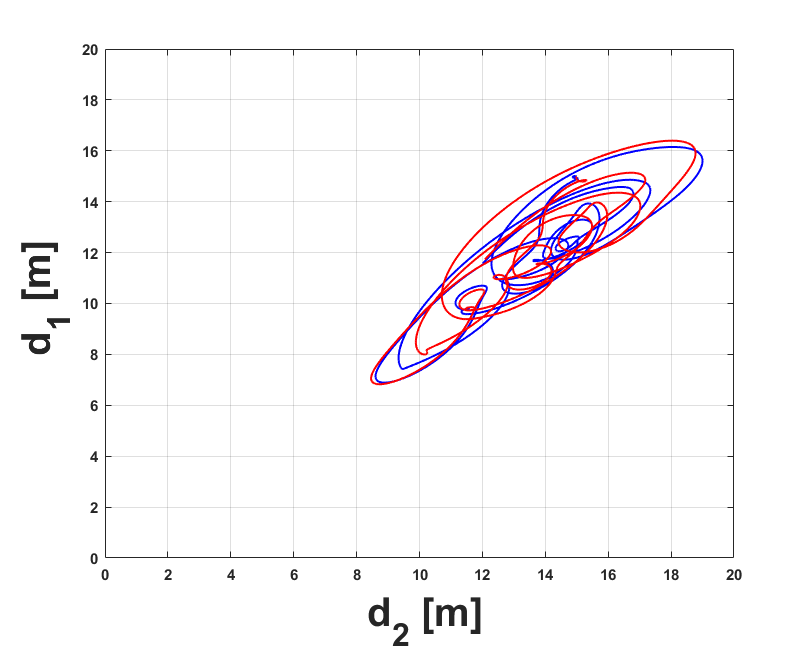}
        \subcaption{Projection onto $d_2$,$d_1$}
        \label{fig:prob1_6_2}
    \end{minipage}
		    \begin{minipage}{.31\textwidth}
        \centering
				\includegraphics[width=1\linewidth, height=0.17\textheight]{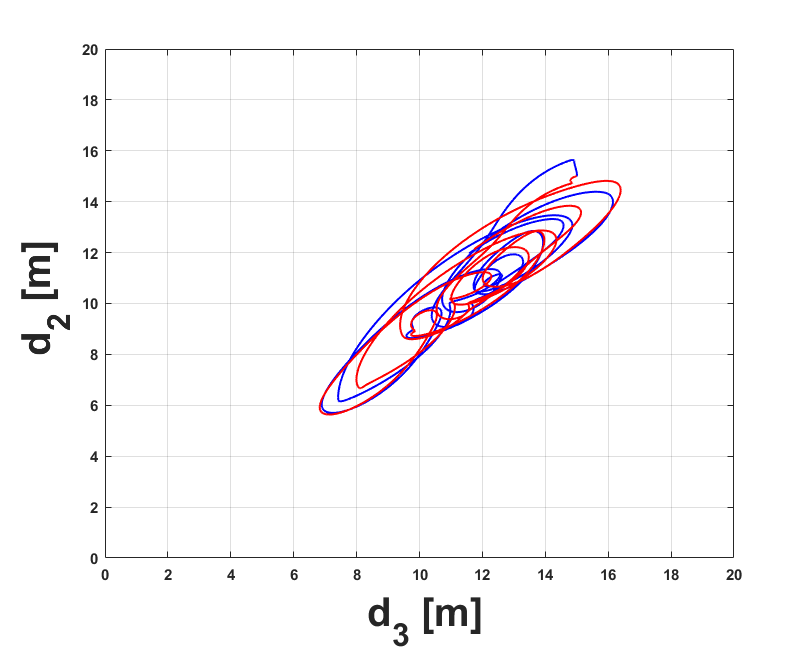}
        \subcaption{Projection onto $d_3$,$d_2$}
        \label{fig:prob1_6_2}
    \end{minipage}
		\begin{minipage}{.31\textwidth}
        \centering
				\includegraphics[width=1\linewidth, height=0.17\textheight]{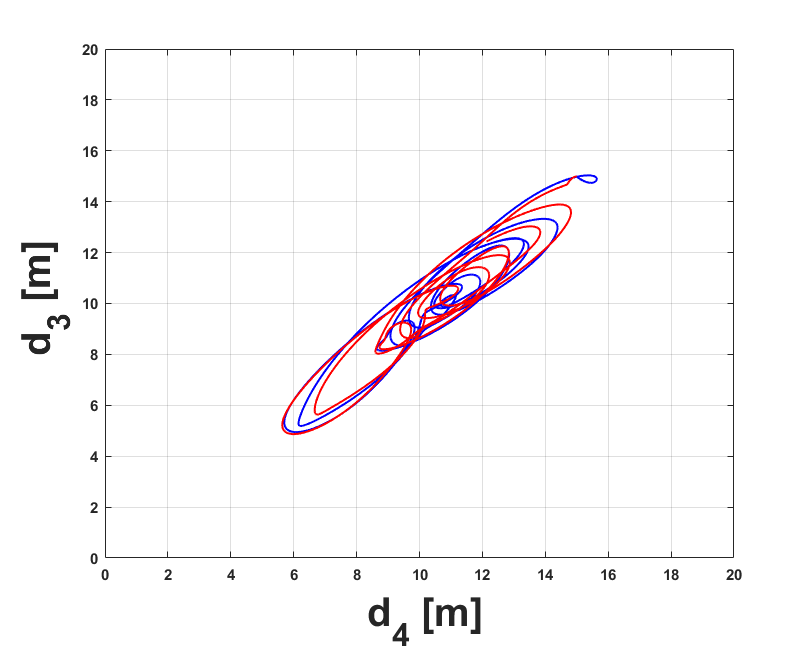}
        \subcaption{Projection onto $d_4$,$d_3$}
        \label{fig:prob1_6_2}
    \end{minipage}		
\\
\begin{minipage}{.31\textwidth}
        \centering
				\includegraphics[width=1\linewidth, height=0.17\textheight]{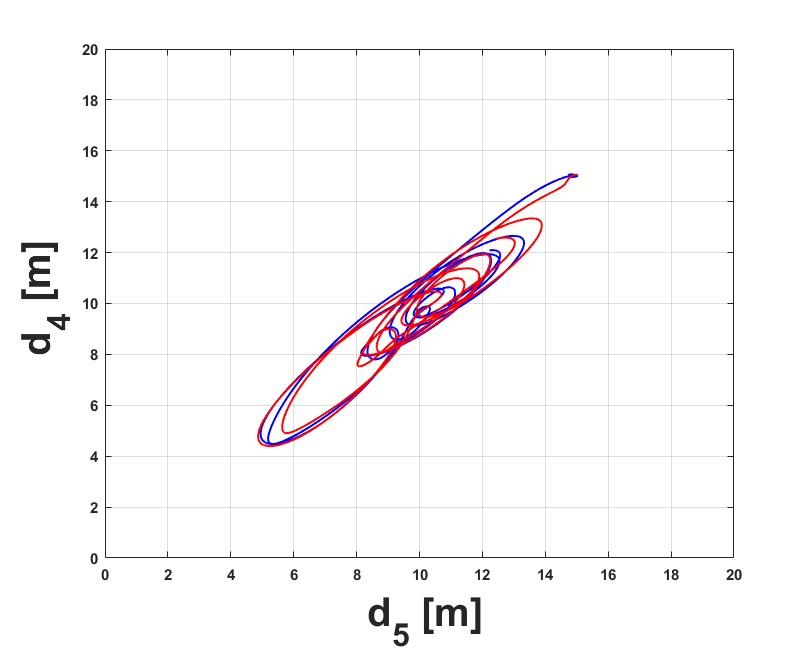}
        \subcaption{Projection onto $d_5$,$d_4$}
        \label{fig:prob1_6_2}
    \end{minipage}
		    \begin{minipage}{.31\textwidth}
        \centering
				\includegraphics[width=1\linewidth, height=0.17\textheight]{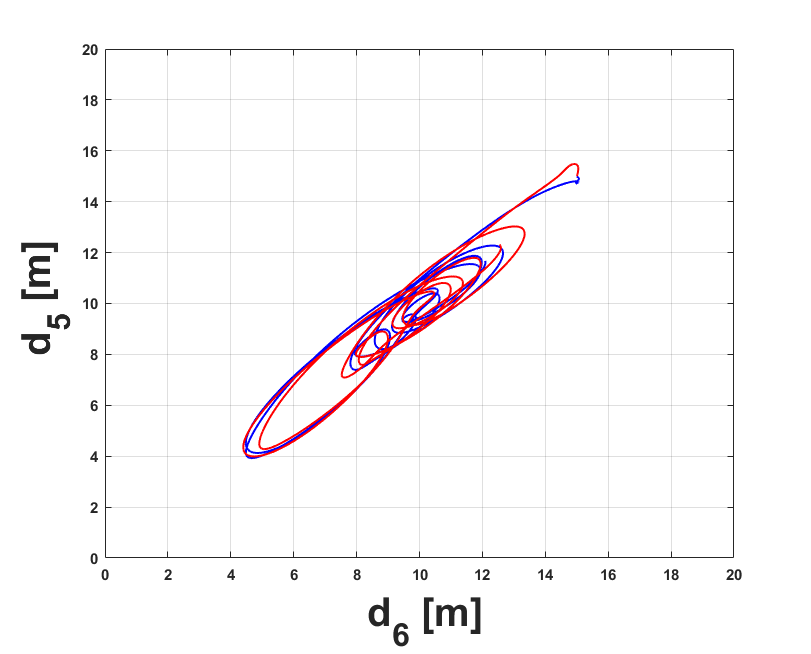}
        \subcaption{Projection onto $d_6$,$d_5$}
        \label{fig:prob1_6_2}
    \end{minipage}
		\begin{minipage}{.31\textwidth}
        \centering
				\includegraphics[width=1\linewidth, height=0.17\textheight]{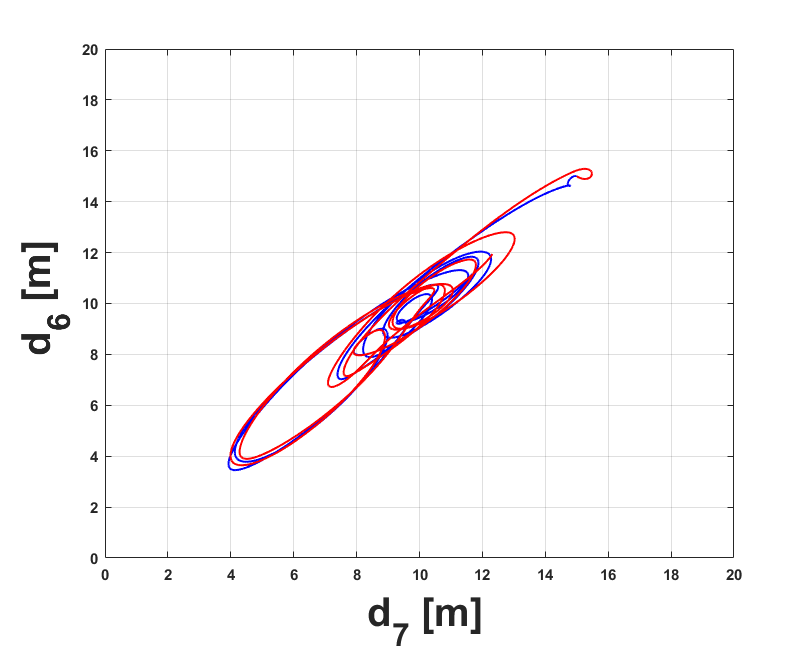}
        \subcaption{Projection onto $d_7$,$d_6$}
        \label{fig:prob1_6_2}
    \end{minipage}		
\\
\begin{minipage}{.31\textwidth}
        \centering
				\includegraphics[width=1\linewidth, height=0.17\textheight]{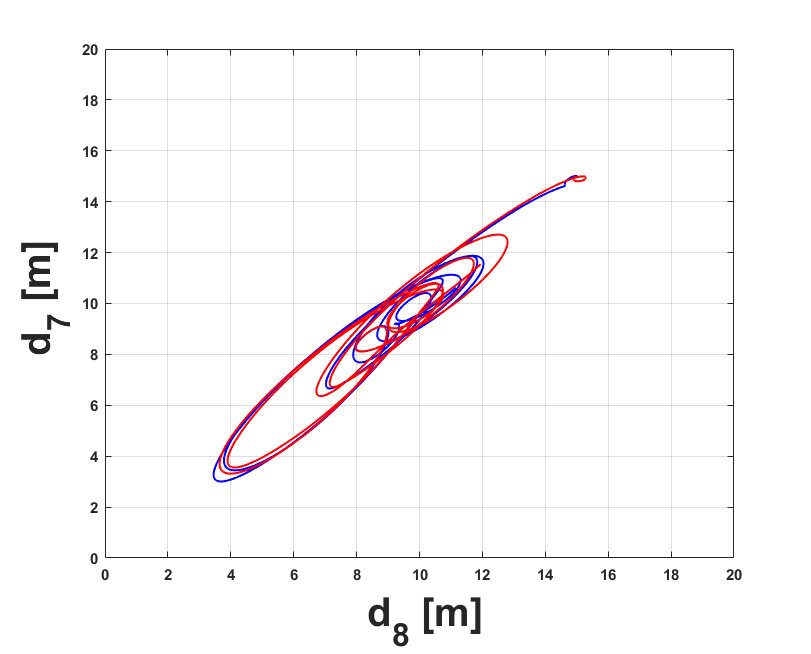}
        \subcaption{Projection onto $d_8$,$d_7$}
        \label{fig:prob1_6_2}
    \end{minipage}
		    \begin{minipage}{.31\textwidth}
        \centering
				\includegraphics[width=1\linewidth, height=0.17\textheight]{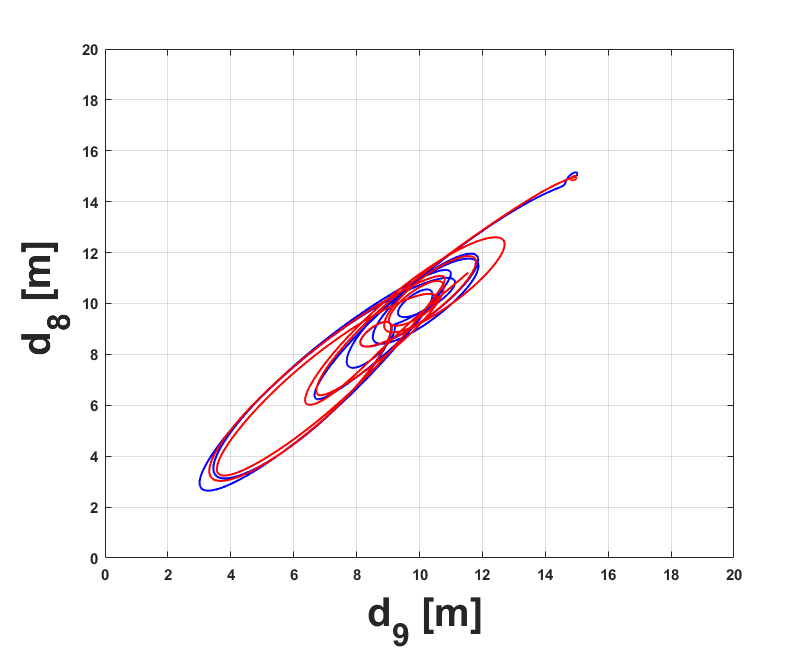}
        \subcaption{Projection onto $d_9$,$d_8$}
        \label{fig:prob1_6_2}
    \end{minipage}
		\begin{minipage}{.31\textwidth}
        \centering
				\includegraphics[width=1\linewidth, height=0.17\textheight]{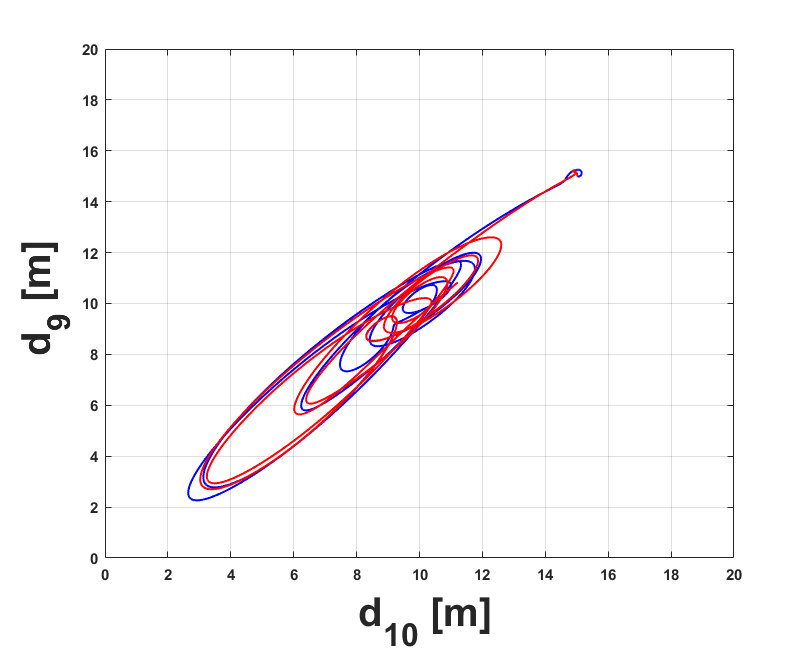}
        \subcaption{Projection onto $d_{10}$,$d_9$}
        \label{fig:prob1_6_2}
    \end{minipage}				
\caption{Reachable inter-vehicle distance states in the platoon}
\squeezeup
\label{Reach}
\end{figure} 
As a result, in (\ref{equ18}) the attacker's impact on the average SINR is removed (i.e., $\mu_j = \sigma_j^2 = 0$).
Then the model only includes the Rician fading channel impact on the CACC performance. 
Lead vehicle as an excitation input signal, takes the commended acceleration profile shown in Figure~\ref{lead}.

Due to the stochastic nature of Rician fading channel model, for the given excitation input in Figure~\ref{lead}, we conduct $10,000$ simulation runs and record the inter-vehicle distance trajectories in the platoon. 
Therefore, for each inter-vehicle distance in the platoon, $10,000$ trajectories are recorded.
To observer the impact of Rician fading channel on each inter-vehicle distance in the platoon, two trajectories are selected uniformly at random.
  
Each subfigure in Figure~\ref{Reach} shows the reachable states for the two adjacent inter-vehicle distances in the platoon. 
The two selected trajectories are different from each other in each subfigure.
This shows that Rician fading channel affects the reachable inter-vehicle distance states in the platoon and deviates the platoon's behavior from normal operation.
 
The results also show that the difference between the two adjacent inter-vehicle distances is reduced as it goes upstream in the platoon.  
For instance, comparison between Figure~\ref{Reach}(a) and Figure~\ref{Reach}(i) shows that inter-vehicle distance differences between the two pairs of $\bold{d_9}$ and $\bold{d_{10}}$ is smaller than the two pairs of $\bold{d_1}$ and $\bold{d_2}$.
This means that, CACC system can reduce the inter-vehicle distance differences as it goes upstream in the platoon. 
This behavior also implies having an stable platoon. 
   
\begin{figure}[!htb]
    \centering
   \begin{minipage}{.5\textwidth}
        \centering
        \includegraphics[width=1\linewidth, height=0.3\textheight]{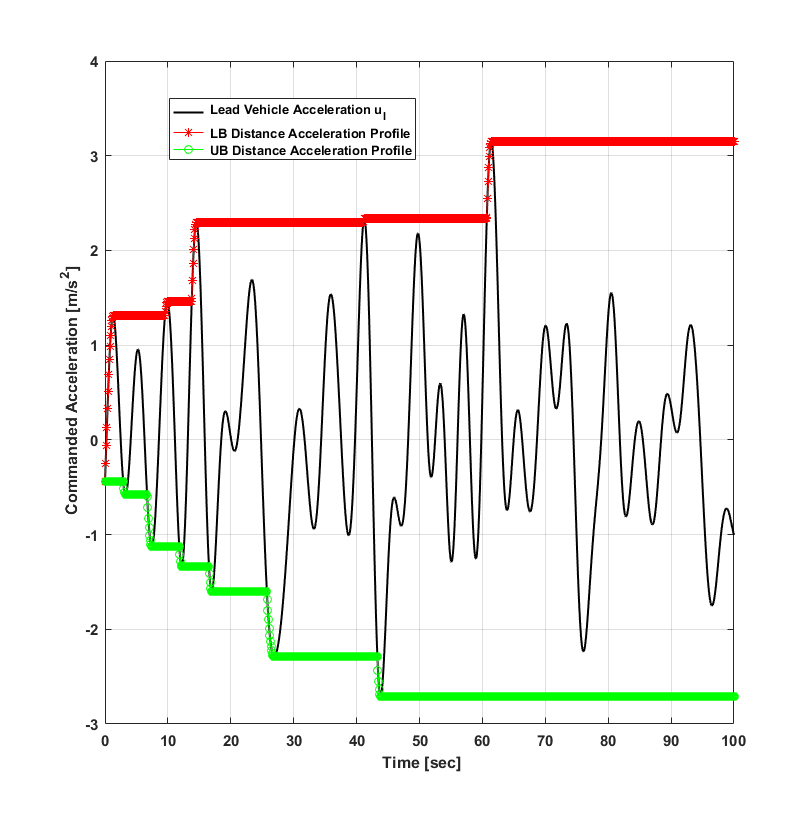}
        \subcaption{Commanded Acceleration}
        \label{a}
    \end{minipage}%
    \begin{minipage}{0.5\textwidth}
       \centering
        \includegraphics[width=1\linewidth, height=0.3\textheight]{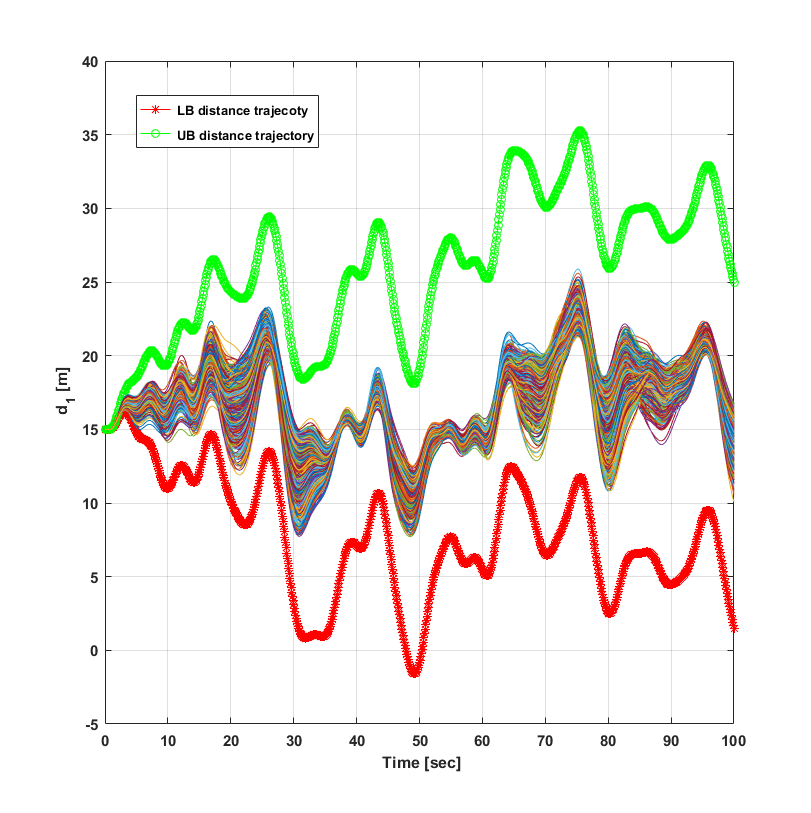}
        \subcaption{Inter-vehicle distance trajectories for Rician fading channel}
        \label{b}
   \end{minipage}
\caption{} 
\squeezeup
\label{Rician}
\end{figure}  

\subsection{Rician Fading Channel and Jamming Attack Impact on the Inter-Vehicle Distance Between Lead Vehicle and its Follower} 
In this subsection, we study the reachable inter-vehicle distance states between the lead vehicle and its follower ($i=1$) in the presence of both the attacker and Rician fading channel.
In fact, we aim to evaluate the performance of the CACC system under Rician fading channel and jamming attacks from the safety perspective, considering coupling impact of the cyber and physical states, as formulated in ~(\ref{equ18},\ref{equ20} and \ref{equ24}).
To do so, for the stochastic dynamical system derived in ~(\ref{equ24}), stochastic reachability analysis as a safety verification method is employed by conducting extensive simulations. 

First, we derive the upper and lower bound of the trajectories of the inter-vehicle distance between the lead vehicle and its follower as follows:

Stochastic random variable $\bold{d_1}$ is given by,

\begin{equation}
\begin{aligned}
\label{equ27}
\bold{d_1}[k+1] = f(\bold{d_1}[k],\tilde u_{0}[k] , u_l[k]) 
\end{aligned}
\end{equation}
where $f$ is a deterministic function defined by the time-invariant matrices listed in ~(\ref{equ10}), and

\begin{equation}
\begin{aligned}
\label{equ28}
\tilde u_{0}[k]=\beta_0^k u_l[k] + \sum_{m = 1}^{k} \beta_0^{m-1} u_l[m-1] \prod_{j=m}^{k}(1-\beta_{0}^{j}) 
\end{aligned}
\end{equation}   

There are several factors that define the inter-vehicle distance states evolution including initial states $\bold{d_1}[0]$ and $u_l[0]$, packets delivery probabilities which according to (\ref{equ18}, \ref{equ19} and \ref{equ20}) depend on $\bold{d_1}[k]$, and the lead vehicle's commanded acceleration profile ($u_l$).
In addition, Equations~(\ref{equ27} and \ref{equ28}) show that the distance value at time $k+1$ also depends on the entire history of the distance value and the lead vehicle commanded acceleration ($u_l$). 
From (\ref{equ27} and \ref{equ28}), maximum value of $u_l$ from time $0$ till time $k$ determines the lower bound value of inter-vehicle distance at time $k+1$.
Similarly, minimum value of $u_l$ from time $0$ till time $k$ determines the upper bound value of the inter-vehicle distance value at time $k+1$.
Note that $u_l$ is a deterministic acceleration profile taken by the lead vehicle.

Therefore, the reachable inter-vehicle distance states set $D_1$ at time $k+1$ is defined as follows:

\begin{equation} 
\begin{aligned}
\label{equ222}
d_1^k \bigg|\tilde u_l^j = max \left\{ u_{l}^{0-j} \text{for} \quad j =1,...,k \right\} \leq Reach\left\{D_1^{k+1}\right\} \leq d_1^k \bigg| \tilde u_l^j = min \left\{ u_{l}^{0-j} \text{for} \quad j =1,...,k \right\} 
\end{aligned}
\end{equation}
 
Figure~\ref{Rician}(a) shows the lead vehicle's commanded acceleration ($u_l$),  and its corresponding two other acceleration profiles that generate upper and lower bound for the inter-vehicle distance trajectories. 
The upper and lower bound distance trajectories for the lead vehicle's commanded acceleration in Figure~\ref{Rician}(a), are shown in Figure~\ref{Rician}(b).

\subsubsection{\textbf{First Scenario}} 
We study the impact of Rician fading channel without the presence of the attacker ($\mu_j=0, \sigma_j^2=0$).  
Figure~\ref{Rician}(b) shows $10,000$ possible inter-vehicle distance trajectories which are upper and lower bounded by the corresponding commanded acceleration shown in Figure~\ref{Rician}(a).
\begin{figure}[!htb]
    \centering
   \begin{minipage}{.5\textwidth}
        \centering
        \includegraphics[width=1\linewidth, height=0.3\textheight]{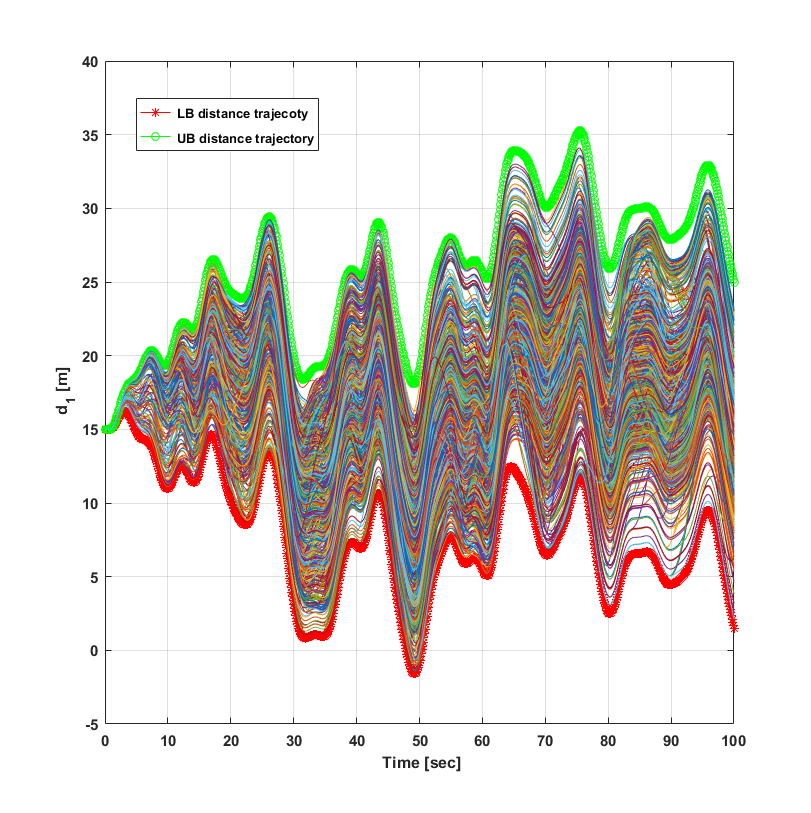}
        \subcaption{Inter-vehicle distance trajectories for Rician fading channel and attacking in whole time horizon}
        \label{a}
    \end{minipage}%
    \begin{minipage}{0.5\textwidth} 
       \centering
        \includegraphics[width=1\linewidth, height=0.3\textheight]{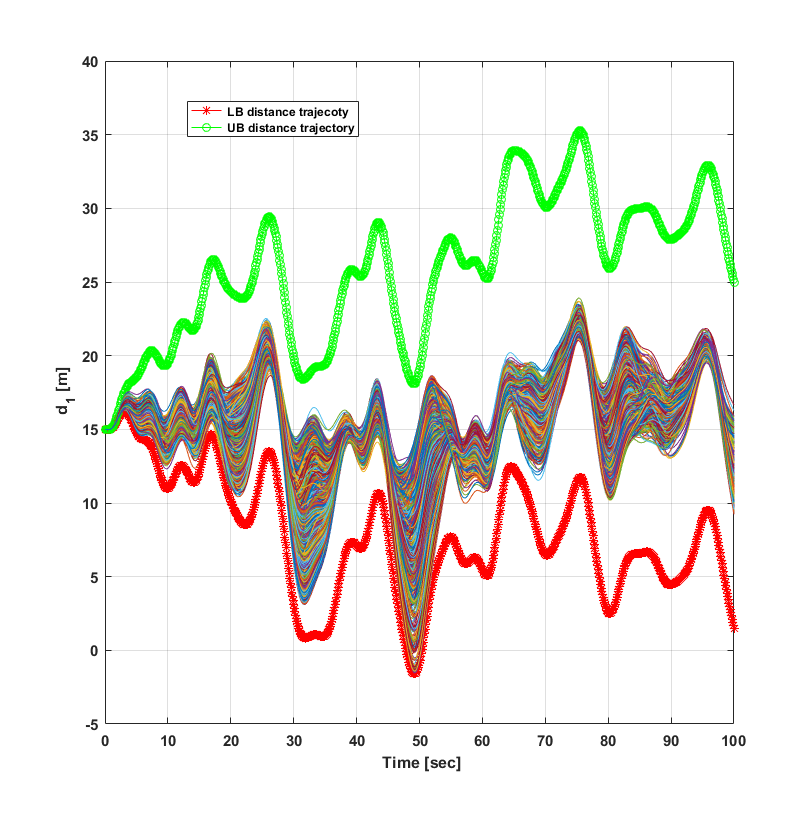}
        \subcaption{Inter-vehicle distance trajectories for Rician fading channel and attacking when the lead vehicle decelerates}
        \label{b}
   \end{minipage}
	\caption{Jamming attack impact} 
\squeezeup 
\label{attack}
\end{figure}
We see that, although the lower bound distance trajectory hits the zero (unsafe state), inter-vehicle distance trajectories are relatively overlapping with each other and the following vehicle maintains a safe distance from the lead vehicle. 

\subsubsection{\textbf{Second Scenario}}
 
In this scenario, we assume that in addition to the Racian fading channel assumption, the jammer also emites its jamming signal over the wireless channel established between the lead vehicle and its follower. 
Figure~\ref{attack}(a) illustrates $10,000$ possible inter-vehicle distance trajectories when the attacker jams the signal for the whole time horizon of $100$ seconds. 
As shown in Figure~\ref{attack}(a), distance trajectories almost cover all the distance states set between the lower and upper bound and some trajectories hit the zero distance (unsafe states).

\subsubsection{\textbf{Third Scenario}} 

In this part, we describe an attacking strategy that is occurring only partially in the time horizon.
From the safety perspective this attacker makes the same impact on the distance trajectories as attacking in the whole time horizon.

To find the best time to attack, we use the fact that according to the velocity-dependent spacing policy, when the lead vehicle decelerates/accelerates,  the inter-vehicle distance decreases/increases as well.
Therefore, intuitively from the safety point of view, the best times for the attacker to launch its jamming signal are the times at which the lead vehicle decelerates.  

Thus, to implement this attacking strategy, we assume that the attacker knows when the lead vehicle decelerates.
This capability for the attacker is obtained by considering that the drone also has been equipped with ACC system with velocity-dependent spacing policy.
Whenever the ACC system senses that the drone's distance from the lead vehicle decreases, it means that the lead vehicle decelerates.
Therefore, the drone's ACC system triggers the jammer signal when it finds that the lead vehicle decelerates.
 
We also assume that, at the times that the attacker is silent (i.e., $\mu_j=0, \sigma_j^2=0$, this happens when the lead vehicle accelerates), wireless channel still suffers from Rician fading.

Figure~\ref{attack}(b) shows the $10,000$ distance trajectories for the proposed attacking strategy. 
As the results show, distance trajectories are dense and closer to the lower bound to make the safety critical situations instead of being distributed within the bound and being closer to the upper bound.
Some trajectories also hit the zero distance (unsafe state).
Also, at the moments when there is no jamming signal (when the lead vehicle accelerates), only the impact of Rician fading channel appears in the inter-vehicle distance trajectories.
Finally, comparing the distance trajectories in Figure~\ref{attack}(a) (attacking within the whole time horizon) and Figure~\ref{attack}(b) (attacking at the times the lead vehicle decelerates), the results show that the partially attacking strategy is effective from the attacker's perspective to make the safety critical situations.

\section{Conclusions and Future Work}
\label{Conclusions} 
In this paper, we have modeled the coupling between cyber (wireless communication) and physical states (inter-vehicle distances) in the vehicle platooning with CACC system and studied the jamming attack impact on the platoon stability and safety in the presence of Racian fading in the channels.
Based on our analysis, we identified that the best location to launch the jamming attack to destabilize the platoon is above the second vehicle in the platoon.
As the attacker moves upstream in the platoon, its impact in terms of destabilizing the platoon is diminished.  
Considering various settings for the headway-time in the CACC control structure, we studied the impact of vehicles' and jammer's transmission signal powers on the mean string stability.
We then analyzed the inter-vehicle distance trajectories between the lead vehicle and its follower in the platoon from the safety perspective when the wireless channel is under jamming attacks.
We computed the upper and lower bound for the first inter-vehicle distance trajectories in the platoon. 
Our analysis show the jamming attacks are more effective in terms of pushing the inter-vehicle distance trajectories to the unsafe states when the lead vehicle decelerates. 

As of future work, we believe that our study highlights many research directions in this area. 
How to defend against a jamming attacker in a real-time manner such that the platoon maintains the stability?
Studying upper and lower bound inter-vehicle distance trajectories for all the inter-vehicle distances in the platoon will be a promising direction.
One other interesting problem is to study the cyber-physical co-attacks.
The attacker can jam the wireless communication while cooperating with a malicious vehicle in the platoon that does not follow the CACC rules and takes disturbing acceleration commands.
Cyber-physical co-defense strategies to avoid safety critical situation is also an interesting research direction.




\bibliographystyle{unsrt}
\bibliography{TCPS-bibliography}

%







\end{document}